\documentclass[12pt,preprint]{aastex}
\usepackage{amsmath}
\usepackage{amssymb}
\usepackage{graphicx}
\usepackage{subfig}
\usepackage{rotating}
\usepackage{booktabs}
\usepackage{threeparttable}
\usepackage{longtable}
\usepackage{lscape}
\usepackage{epsfig}

\slugcomment{July 13th  2011}

\def\msun{M_\odot}
\def\mbh{M_{\rm{BH}}}

\def\<{\,\langle\langle}
\def\>{\,\rangle\rangle}

\author{Yan-Fei Jiang\altaffilmark{1}, Jenny E. Greene\altaffilmark{2}, Luis C. Ho\altaffilmark{3} , Ting Xiao\altaffilmark{4} \& Aaron J. Barth\altaffilmark{5}  }
\affil{$^1$Department of Astrophysical Sciences, Princeton
University, Princeton, NJ 08544, USA}
\affil{$^2$Department of Astronomy, University of Texas at Austin, 1 University Station, C1400, Austin, Texas 
78712, USA}
\affil{$^3$The Observatories of the Carnegie Institution for Science, 813 Santa Barbara 
Street, Pasadena, CA 91101, USA}
\affil{$^4$Key Laboratory for Research in Galaxies and Cosmology,
Department of Astronomy,
University of Science and Techonology of China, 
Chinese Academy of Sciences, Hefei, Anhui 230026, China
}
\affil{$^5$Department of Physics and Astronomy, University of
California at Irvine, Irvine, CA 92697, USA}

\begin{document}

\title{The Host Galaxies of Low-Mass Black
  Holes\footnote{ Based on observations made with the NASA/ESA
    Hubble Space Telescope, obtained at the Space Telescope Science
    Institute, which is operated by the Association of Universities
    for Research in Astronomy, Inc., under NASA contract NAS
    5-26555. These observations are associated with program
    GO-11130.}}

\begin{abstract}
  Using \emph{HST} observations of 147 host galaxies of low-mass black
  holes (BHs), we systematically study the structures and scaling
  relations of these active galaxies. Our sample is selected to have
  central BHs with virial masses $\sim10^5-10^6\msun$. The host
  galaxies have total $I$-band magnitudes of $-23.2<M_I<-18.8$ mag 
  and bulge magnitudes
  of $-22.9<M_I<-16.1$ mag. Detailed bulge-disk-bar decompositions with GALFIT
  show that $93\%$ of the galaxies have extended disks, $39\%$ have
  bars and $5\%$ have no bulges at all at the limits of our
  observations.  Based on the S\'ersic index and bulge-to-total ratio,
  we conclude that the majority of the galaxies with disks are likely
  to contain pseudobulges and very few of these low-mass BHs live in
  classical bulges.  The fundamental plane of our sample is offset
  from classical bulges and ellipticals in a way that is consistent
  with the scaling relations of pseudobulges.  The sample has smaller
  velocity dispersion at fixed luminosity in the Faber-Jackson plane,
  compared with classical bulges and elliptical galaxies. The galaxies
  without disks are structurally more similar to spheroidals than to
  classical bulges according to their positions in the fundamental
  plane, especially the Faber-Jackson projection.  Overall, we suggest
  that BHs with mass $\lesssim 10^6 \msun$ live in galaxies that have
  evolved secularly over the majority of their history.  A classical
  bulge is not a prerequisite to host a black hole.
\end{abstract}


\section{Introduction}
\label{sec:intro}

In the past decade, we have found strong correlations between
supermassive black hole (BH) masses ($\sim 10^6-10^9\msun$) and the
properties of host bulges (e.g, the $M_{\rm BH}-\sigma_{\ast}$
relation; \citealt{Tremaineetal2002}, and the $M_{\rm BH}-L_{\rm
  bulge}$ relation; \citealt{MarconiHunt2003}). However, very little
is known about BH-bulge correlations for low-mass BHs ($<10^6\msun$)
or late-type galaxies. The correlations between BH mass and galaxy
properties in this low-mass regime provide important constraints on
the formation mechanisms of the first primordial seed BHs
\citep[e.g.,][]{volonterinatarajan2009}.  Furthermore, low-mass BHs 
are expected to be a major source of gravitational
radiation \citep[e.g.,][]{hughes2002}.  Given the strong correlations
between bulge properties and BH mass, the question remains whether
supermassive BHs can exist in bulgeless galaxies, and if so how
massive they become.

It was originally thought that BH mass was linked exclusively with
bulge mass. For instance, the nearby late-type spiral galaxy M33 does
not contain a central BH more massive than 1500 $\msun$
\citep[e.g.,][]{Gebhardtetal2001,Merrittetal2001}. Similarly, a
central BH in the nearby dwarf galaxy NGC 205 has an upper limit of
$2.2\times10^4$ $\msun$ on its mass
\citep[e.g.,][]{Vallurietal2005}. On the other hand, recent observations 
show that central massive BHs can also exist without classical bulges
\citep[e.g.,][]{FilippenkoHo2003,GreeneHo2004,Barthetal2004,
  Greenetal2007,Satyapaletal2007,Shieldsetal2008,Satyapaletal2009,
  Barthetal2009}.  Prior to our work, NGC 4395 and POX 52 were the
only galaxies known to host BHs with $\mbh\lesssim 10^6~\msun$.  NGC
4395 is an Sdm spiral with no bulge.  POX 52 is a spheroidal galaxy,
also sometimes called a dwarf elliptical, although we follow the
naming convention of \citet{Kormendyetal2009}.  One has a disk, and
the other has no disk.  Neither has a classical bulge component.  A
larger sample of low-mass BHs is needed to understand the properties
of their host galaxies.

Finding low-mass BHs is particularly challenging.  Unlike massive BHs
in nearby galaxies, it is nearly impossible to measure dynamical BH
masses for BHs with $\lesssim10^6\msun$ outside of the Local Group.
We cannot yet resolve their gravitational spheres of influence,
although it is possible to place interesting constraints on central BH
masses in nearby objects \citep[e.g.,][]{Barthetal2009,Sethetal2010}.
Instead, we rely on indirect methods to estimate the ``virial'' masses
of actively accreting BHs.  Based on the broad H$\alpha$ profile and
the calibrated radius-luminosity relation
\citep[e.g.,][]{Bentzetal2009}, \cite{GreeneHo2004} presented 19
galaxies with virial BH masses $\lesssim10^6\msun$.  The sample has
since increased to 174 galaxies (\citealt{Greenetal2007}, see also
\citealt{Dongetal2007}).  From the SDSS data alone, we do not
learn much about the host galaxy properties.  Here we present a
study of the host galaxies of this large sample of low-mass BHs using
\emph{Hubble Space Telescope} (\emph{HST}) observations.  Our primary
goal is to determine the morphological types of galaxies hosting
low-mass BHs, in particular the types of bulges that they contain.
From the SDSS images, we know that the galaxies are $\sim 1$ mag below
$L^*$ \citep{GreeneHo2004}.  In this luminosity range, galaxies have
many different morphologies, ranging from small elliptical galaxies
(e.g., M32) to late-type spirals and spheroidals. We want to determine
whether a bulge is a necessary requirement to host a central
supermassive BH.

First, we will measure the fraction of low-mass BHs without a
bulge-like component of any kind. Bulgeless galaxies observed in the
nearby universe, such as NGC 4395 and NGC 6946 (e.g.,
\citealt{FilippenkoHo2003,Shihetal2003,Boomsmaetal2008}), are
recognized as a challenge to the cold dark matter galaxy formation
scenario
\citep[e.g.,][]{KormendyFisher2008,Kormendyetal2010,PeeblesNusser2010}.
The fact that there are BHs in bulgeless galaxies implies that a bulge
is not a necessary condition for the formation of BHs.
\cite{Satyapaletal2009} study $18$ truly bulgeless Sd/Sdm galaxies
with {\it Spitzer} and find only one active galaxy (NGC 4178), which
suggests that BHs in bulgeless galaxies may be truly rare.  X-ray
observations find active galactic nuclei (AGNs) in $\sim 25\%$ of
Scd---Sm galaxies (\citealt{DesrochesHo2009}).  
Larger samples of such bulgeless host galaxies will
elucidate their nature.

Second, we will determine what fraction of the low-mass BH hosts have
no disk.  These may be either elliptical or spheroidal galaxies
\citep[e.g.,][]{Ferrareseetal2006,Kormendyetal2009}.  More massive 
BHs are found in elliptical galaxies, but these lower-mass systems have 
stellar masses that are consistent with being either small ellipticals or 
spheroidals.  We will use their scaling relations (e.g., the fundamental plane) 
to distinguish between these two types.  The differing structures of 
spheroidal and elliptical galaxies likely reflects different formation 
histories \citep[e.g.,][]{Kormendyetal2009}, and so the question is whether BH 
formation and growth occurs in spheroidal systems.

Third, for the disk galaxies with a bulge component, we ask whether they are
classical bulges or not.  Observational evidence is building that
there are two different kinds of bulges, namely classical bulges and
pseudobulges (e.g., \citealt{Kormendyetal2004}).  Pseudobulges have
properties, including rotational support, exponential profiles, and
ongoing star formation, that implicate the importance of secular
processes such as bar transport in their build-up. We will measure the
structure of the bulges from the \emph{HST} observations to decide whether
they are pseudobulges or classical bulges.

Finally, many recent papers have hinted that the scaling relations
between low-mass BHs and their bulges are systematically different
compared with more massive BHs \citep[e.g.,][]{Hu2008,Greenetal2008,
  GadottiKauffmann2009, Greenetal2010}.  Ultimately, we will use the
structural measurements presented here to examine the
$M_{\rm BH}-L_{\rm bulge}$ and $M_{\rm BH}-M_{\rm bulge}$ correlations 
for this sample \citep{Jiangetal2011}.

In \S\ref{sec:data}, we describe the data and reduction process.  In
\S\ref{sec:image}, we present the image decompositions.  We explain
how the uncertainties and upper limits are estimated in
\S\ref{sec:uncertainty}.  Morphological results are given in
\S\ref{sec:morphology} and scaling relations are given in
\S\ref{sec:pseudobulge}. Finally, in \S\ref{sec:summary}, we summarize
the paper.  The following cosmological parameters have been adopted:
$H_0=100h=71$ km s$^{-1}$ Mpc$^{-1}$, $\Omega_m=0.27$ and
$\Omega_{\Lambda}=0.75$ (\citealt{Spergeletal2003}).  Galactic
extinction is calculated based on the fitting formula given by
\cite{Cardellietal1989}.

\section{The Data}
\label{sec:data}
In this section, we first briefly present the sample selection. We then describe 
the data reduction processes used to produce the final images for analysis.

\subsection{Our Sample}

The galaxies presented here are drawn from the sample described in
\cite{Greenetal2007}.  This sample is selected from all broad-line
active galaxies in the Fourth Data Release \citep[DR4,][]{SDSS4} of the Sloan
Digital Sky Survey \citep[e.g.,][]{Yorketal2000,Greenetal2007a} with $z<0.35$.  
First, the DR4 spectra are continuum-subtracted using a principal component 
analysis developed by \cite{Haoetal2005}.  Then objects
with high rms deviations above the continuum in the broad
H$\alpha$ region are selected and more detailed profile fitting is
applied to isolate those objects with H$\alpha$ profiles that are broad compared to
the narrow [\ion{S}{2}] and [\ion{N}{2}] lines.  

Virial BH masses are estimated for all targets, using the
broad-line region (BLR) gas as the dynamical tracer.  The virial mass is
simply $\mbh=fR(\Delta v)^2/G$, where $R$ is the size of the BLR, 
$\Delta v$ is a measure of the broad-line width, such
as full-width at half-maximum (FWHM), and $f$ is a dimensionless factor
that accounts for the unknown geometry and kinematics of the
BLR.  A few dozen AGNs have direct
measurements of their BLR sizes from reverberation mapping,
a measurement of the lag between continuum and line variations
(e.g., \citealt{Petersonetal2004,Bentzetal2009,Denneyetal2010}).  
An empirical correlation between BLR radius and AGN luminosity 
(the radius-luminosity relation) is then used to infer the BLR sizes for other AGNs 
(e.g., \citealt{Bentzetal2006, Bentzetal2009}).  In this work, we use the luminosity 
of H$\alpha$\ to infer the BLR radius \citep{GreeneHo2005}.  The virial mass is then
estimated from the luminosity and FWHM of H$\alpha$ as (e.g.,
\citealt{Greenetal2007}; \citealt{Wooetal2010})
\begin{equation}
\mbh=3.0\times10^6\left(\frac{L_{\rm{H}\alpha}}{10^{42}\
\text{ergs}\ \text{s}^{-1}}\right)^{0.45}
\left(\frac{\text{FWHM}_{\text{H}\alpha}}{10^3\  \text{km}\
\rm{s}^{-1}}\right)^{2.06}\msun .
\end{equation}
Since we cannot yet determine the value of $f$ for each AGN
appropriately, we use a single value of $f=0.75$ (e.g.,
\citealt{Kaspietal2000}) that is intended to represent an ensemble
average over our sample.  The virial masses we use in this paper are
calculated according to the above formula with the H$\alpha$
luminosity and FWHM based either on the SDSS spectrum or a
higher-resolution spectrum from ESI on Keck \citep{Sheinisetal2002} or MagE on
Magellan (\citealt{Marshalletal2008}). The spectra are presented in
\cite{Xiaoetal2011}.

From this parent sample, the final 174 BHs were selected
to have virial masses smaller\footnote{After improvement of the BH
  mass estimator, a few of the BHs have masses that are larger than this value.}
than $2\times10^6\msun$ (\citealt{Greenetal2007}).  An additional 
55 galaxies were presented with far less certain broad-line masses 
(called `c'). 
 The \emph{HST} snapshot pool was taken from these 229 galaxies.
These galaxies have a median g-band magnitude of $M_g=-19.3$ and a median 
color of $\langle g-r \rangle =0.7$ magnitude. 
The median redshift is $\langle z \rangle = 0.085$, with a maximum redshift of
$z = 0.35$. They have BHs with virial masses ranging from
$M_{\rm BH} = 6.2\times10^4\msun$ to $3.8\times10^7\msun$ with a median value
$1.2\times10^6\msun$.
The BHs are radiating at high fractions
of their Eddington limits and most are radio-quiet.
More properties 
of these galaxies are described in \cite{Greenetal2007}.

\begin{sidewaystable}
\centering
\begin{threeparttable}
\caption{Observations Summary} \label{observations}
\begin{tabular}{ c  c  c c c  c c }
\hline \hline
              &                         &        & $A_I$ &                		& 	 scale   		&  $\log\mbh$	\\
Name & SDSS Name  & $z$ & (mag) & observation Date &    (arcsec/kpc)        &	($\msun$)		\\
 (1)        &              (2)      &   (3) & (4)      &   (5)                           &      (6)     		&	(7)		\\
 \hline
$0022- 0058$ & SDSSJ002228.36$-$005830.6 &    0.106  &    0.32    &    $2008-08-22$      &    0.42   &  5.7	\\
$0024 -1038$ & SDSSJ002452.53$-$103819.6 &    0.103  &    0.15    &   $2008-09-25  $     &    0.43   & 6.2	\\
$0117 -1001$ & SDSSJ011749.81$-$100114.5 &    0.141  &    0.50    &    $2008-08-19 $     &    0.31   & 5.8	\\
$0120 -0849$ & SDSSJ012055.92$-$084945.4 &    0.125  &    0.14    &    $2008-06-09  $    &    0.35  &  6.3	\\
$0158 -0052$ & SDSSJ015804.75$-$005221.8 &    0.0804  &    0.05    &  $2008-09-27  $      &  0.56    & 5.9	 \\
 $0228 -0901$ & SDSSJ022849.51$-$090153.7 &    0.0722  &    0.13    &   $2007-10-22  $     &  0.63    & 5.4	 \\
 $0233 -0748$ & SDSSJ023310.79$-$074813.3 &    0.0310  &    0.20    &   $2008-11-01 $      &  1.52   &  6.0	 \\
 $0240+0103$ & SDSSJ024009.10$+$010334.5 &   0.196  &    0.038    &   $2007-11-19      $ &   0.22    &  5.8	\\
 $0304+0028$ & SDSSJ030417.78$+$002827.3 &   0.0444  &    0.071    &   $2008-09-23$       &  1.05    &6.1	 \\
 $0325+0034$ & SDSSJ032515.59$+$003408.4 &   0.102  &   0.071    &    $2008-11-25 $     &   0.44   &  6.0	 \\
 $0327- 0756$ & SDSSJ032707.32$-$075639.3 &   0.154  &   0.15    &   $2008-12-05    $   &    0.28   &   5.7	\\
 $0347 + 0057$ & SDSSJ034745.41$+$005737.2 &   0.179  &   0.061    &   $ 2008-07-27  $    &  0.24    & 7.6	 \\
 $0731+ 3926$ & SDSSJ073106.86$+$392644.6 &   0.0483  &   0.057    &  $2008-11-22    $    & 0.96     &	 6.1	 \\
 $0735+ 4235$ & SDSSJ073505.65$+$423545.6 &   0.0858  &   0.041    &   $2008-09-09  $     &  0.53    &	6.2	 \\
 $0744+ 2430$ & SDSSJ074423.44$+$243046.3 &   0.117  &   0.072    &  $2008-11-13   $     &  0.38    &   6.2		\\
 $0748+ 4540$ & SDSSJ074810.36$+$454003.1 &  0.143   &   0.15    &  $ 2008-11-17  $     &  0.30     &  6.3		\\
 $0748+ 4052$ & SDSSJ074825.27$+$405217.8 &  0.136   &   0.37    &   $2008-11-23$       &  0.32     &  5.7		\\
 \hline
 \end{tabular}
\begin{tablenotes}
\item [1]   Table \ref{observations} is published in its entirety in the electronic edition of the {\it Astrophysical Journal}. 
A portion is shown here for guidance regarding its form and content.
\item [2]    Col (1):    Abridged SDSS name.
\item [3]    Col (2):    Full SDSS name as well as coordinates.
\item [4]    Col (3):    Galaxy redshift.
\item [5]    Col (4):    Galactic extinction in the {\it I}-band.
\item [6]    Col (5):    Date of observation.
\item [7]    Col (6):    Physical scale of image.
\item [8]    Col (7):    Virial BH mass.

\end{tablenotes}
\end{threeparttable}
\end{sidewaystable}

\subsection{Observations}

In order to study the detailed structure of the host galaxies, we
were awarded a snapshot survey with Wide Field Planetary Camera 2
(WFPC2) on \emph{HST} in cycle 16.  
A total of 147 galaxies from \cite{Greenetal2007} 
were observed during this program, including some from the `c'
sample.  Each galaxy was placed at the center of the Planetary Camera
CCD. The WFPC$2$
field-of-view is divided into four cameras by a four-faceted pyramid
mirror near the \emph{HST} focal
plane\footnote{http://www.stsci.edu/hst/wfpc2}. Each of the cameras
contains an $800\times$ $800$ pixel detector. Three identical cameras, 
with a plate scale of $0\farcs 1$~pixel$^{-1}$, form an ``L" shape.
The fourth camera, the Planetary Camera, is located at the upper right
corner of the ``L" and has a finer plate scale of
$0\farcs046$ per pixel. The total effective field of view is 
$\sim 32\farcs2\times 32\farcs2$.
The typical  FWHM of the PSF for the
Planetary Camera is $\sim1.7$ pixels in the F814W filter.  

Details of the observations are summarized in Table
\ref{observations}.  For each object we obtained a short (30 sec)
exposure in case of saturation, followed by two dithered $\sim 600$
sec exposures with the F814W filter.  The mean wavelength of this
filter is $8269$\AA\ and the central wavelength is $8012$\AA, which is
a little different from the Johnson $I-$band filter. As discussed in
\cite{Greenetal2008}, independent of galaxy color the difference in
magnitude between F814W and $I$ is so small ($< 0.05$ mag) that
throughout we will refer to the \emph{HST} observations as $I$-band
images.

\subsection{Data Reduction}
\label{Sec:Data}

The raw \emph{HST} data must be processed before they
can be used for scientific analysis. Since the short exposure is
only used in two cases (described 
at the end of this section), we focus here on the long exposures. 

The first step is to remove cosmic rays using the 
identification script LACosmic (\citealt{Dokkum2001}), which can
detect cosmic ray hits of arbitrary shape and size. Second, we shift
the second exposure by $11 \times 11$ pixels to undo the dither made
during the observations. Due to charge-transfer inefficiencies in
the WFPC2 CCDs and accumulated radiation damage to the WFPC2 CCDs, the
PC images now contain cosmic-ray trails or ``ghost CRs'' that are not
removed by LACosmic. The ghosts are removed as follows. For each pixel
in one image, we check if it deviates by more than 2 $\sigma$ from the
median in an $11\times11$ pixel box in the other image. If so, it is
likely a ghost CR and it is replaced by the median from the other
image. After the two long exposures are combined properly, the final
image is ready for analysis.

One exception to the general procedures above occurs when 
the image core is saturated. The PC chip
becomes nonlinear when the counts reach $\sim3000$ and saturates at
$\sim4000$ counts. Here we flag a pixel as saturated when the counts exceed 3000.

In our sample, there are only two galaxies with saturated cores,
$\text{SDSS~J}030417.78+002827.3$ and
$\text{SDSS~J}115341.77+461242.2$.  For these galaxies we 
replace the saturated pixels with the scaled pixels from the 30 second
exposure image. We have also checked the radial profiles for the
corrected images to make sure that the radial profiles are smooth and that the
correction is applied appropriately.  The saturated core of the 
stellar PSF is replaced in a similar fashion.

\section{Image Decomposition}
\label{sec:image}

Our analysis follows that of the pilot
sample presented in \cite{Greenetal2008}. We first present
our PSF model and sky determinations.  Then we perform two-dimensional 
image decomposition to study the bulges, disks, and bars of these galaxies.

\subsection{The Point-Spread Function}
\label{sec:PSF} An accurate knowledge of the PSF
is required for image decomposition, but it is particularly
important in our case because the AGN is a bright, unresolved
central source. Based on the fitting described below, the fraction
of light contributed by the AGN relative to the host galaxy varies over a large 
range across the sample, from 0-80\%, with a median of 5\%.  The effective 
radii of the bulge components vary from $0.05\arcsec$ to $6\arcsec$. 
Since our primary goal is to study the galaxy bulges,  
it is important to have an accurate PSF model, especially for the 
faint bulges.

The PC chip of WFPC2 has a small field of view and so we do not have
many bright stars in these images that can be used as a PSF
model. Here we use the Tiny Tim software \citep[][]{Krist1995} to
generate a PSF image for each object.  Given the filter and the
location of the AGN in the camera, Tiny Tim can model both the spatial
and spectral variations in the PSF and account for the effect of
charge diffusion.  For each galaxy, we generate a PSF model with Tiny
Tim at the location of the AGN.  We analyze the image based on this
PSF.

Any bright stars near the center of an image can also serve as an
alternative PSF star.  Of the 147 images, only one contains a bright
star near the center of the image that is significantly brighter than
the AGNs and thus can be used as a PSF model. This star is in the field of
SDSS$~$J084234.50+031930.6 and is only 96 pixels
away from the center of the galaxy.  The core is
saturated, but we replace the core using the short exposure.  This
star is used as an alternative PSF for all the galaxies so that we can
estimate the uncertainty due to the PSF model.

\begin{figure}
\centering
\includegraphics[width=1.0\hsize]{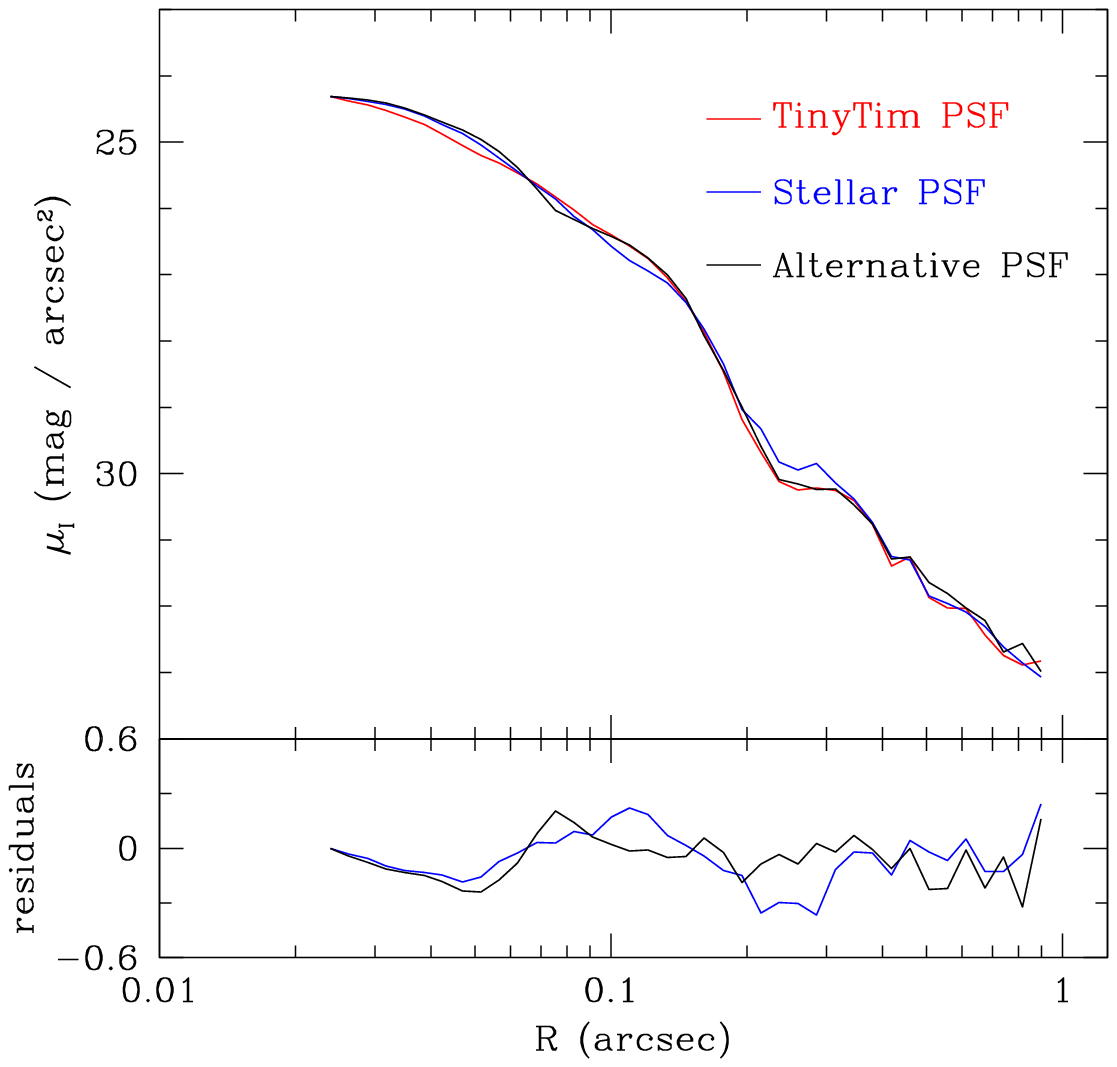}
\vspace{3mm} \caption{Comparison of the radial profiles of three different PSF models.
The red line is the PSF model generated by Tiny Tim for the galaxy
 SDSSJ084234.50+031930.6. The blue line is for the bright
 star in the image of this galaxy. The black line is for a PSF model
 in the WFPC2 PSF library. They are all normalized to an arbitrary  surface
 brightness at $0\farcs02$. }
 \label{comparePSF}
\end{figure}

There are 14 galaxies for which the fitting fails with this alternate
stellar PSF.  In these cases, we select another from the PSF archive
of WFPC2 with the same observational parameters (such as filter) as
our images. This PSF is also used as an alternate PSF to estimate the
uncertainty due to the PSF model, but is only used when the stellar
PSF model fails. In Figure \ref{comparePSF}, we compare the radial
profiles of the three PSF models arbitrarily normalized
at 
$0\farcs02$.  As shown in this plot, the three PSF models are very
similar, although they differ somewhat around $0\farcs1\sim 0\farcs3$. The
difference is due to small spatial variations as shown in the
left-hand panel of Figure 2 in \cite{Kimetal2008}. The spatial
variations for the three PSF models are smaller than $100$ pixels so
that the cores ($\sim$ 2 pixels) are very similar. Thus, there are
small variations between these and the Tiny Tim PSF model that can be
used to quantify the impact of an imperfect PSF model.

\begin{sidewaystable}
\centering
\begin{threeparttable}
\caption{Components of the galaxies} \label{parameters}
\begin{tabular}{ c  c  c c c  c  c  c  c c c c c c }
\hline \hline

  & \multicolumn{2}{c}{Nonparametric} &  & \multicolumn{10}{c}{Parametric}  \\
  \cline{2-3}\cline{5-14}\\
   &   &   &   &  & \multicolumn{2}{c}{Disk}  &  &  \multicolumn{3}{c}{Bulge}  &    &  &    \\
   \cline{6-7}\cline{9-11}\\

  & AGN & Galaxy &  & AGN  &  $r_s$ &   & &  n & $r_e$ &  &   & Bulge/ &  others \\
 No. & $m_I $  &  $m_I $ &  &  $m_I $  &  (kpc)  &  $m_I $  &   &   &  (kpc)  & $m_I$  &   &  Galaxy    &  \\
 (1)  & (2)  & (3)  &   & (4)  &   (5)  & (6)  &  & (7)   &  (8)  &  (9)  & &   (10)    &  (11)   \\
  \hline

$0022-0058$  &   18.7 & 18.3 &  & 19.3$\pm$2.9 & 1.28$\pm$0.19 & 18.43$\pm$0.16  & & 2 & 0.55$\pm$0.43  & 19.40$\pm$0.37  &  & 0.29  & \\

$0024 -1038$ &  18.7 & 17.1 & &   20.6$\pm$1.8 & 1.22$\pm$0.85 & 17.10$\pm$0.01  &  &  2 & 0.14$\pm$0.007  & 18.93$\pm$0.10 &  & 0.16  &  \\

$0117 -1001$ &   19.2 & 18.7 &  & 19.3$\pm$0.4 & 10.76 & $>17.80$   &   & 2,3,4 & 2.36$\pm$0.10  & 18.40$\pm$0.03  &  & 1.00 & \\

 $0120 -0849$ &   19.8 & 18.8 &  &  $>21.2$ & 2.44$\pm$1.90 & 18.95$\pm$0.07  &   & 2,3,4 & 0.31$\pm$0.06  & 19.28$\pm$0.02  & &   0.42 & \\

$0158 -0052$ &   19.5 & 17.7 &  &  20.6$\pm$0.5 & 7.21 & $>17.01$ &  & 3 & 1.58$\pm$0.07  & 17.42$\pm$0.03 &   & 1.00 &  \\

$0228 -0901$ &   19.8 & 17.0 &   & 20.7$\pm$1.3 & 2.15$\pm$0.55 & 17.01$\pm$0.02   & & 2,3,4 & 0.21$\pm$0.08  & 20.35$\pm$0.29 & &  0.04 & \\

$0233 -0748$ &   18.8 & 14.8 &  &  20.8$\pm$0.4 & 1.69$\pm$0.29 & 15.56$\pm$0.10   & & 2 & 0.61$\pm$0.05  & 15.92$\pm$0.09 &   &  0.42 & \\

$0240+0103$ &   21.2 & 19.4 &   & 20.0$\pm$1.1 & 3.39 & $>19.20$ &   & 2,3 & 0.74$\pm$0.33  & 19.80$\pm$0.29 & &  1.00  & \\

$0304+0028$ &   16.9  & 14.9 &  &  17.9$\pm$0.3 & 5.05$\pm$1.20 & 14.74$\pm$0.23   & & 3 & 0.19$\pm$0.11  & 16.68$\pm$0.06 &  &  0.12   & bar\\

\hline

\end{tabular}
\begin{tablenotes}
\item     Table \ref{parameters} is published in its entirety in the electronic edition of the {\it Astrophysical Journal}. A portion is shown here for
guidance regarding its form and content.
\item Col (1):    Abridged SDSS name for this galaxy.
\item Col (2):    Nonparametric magnitude (mag) for the AGN component.
\item Col (3):    Nonparametric total magnitude (mag) for the host galaxy.
\item Col (4):    Parametric magnitude (mag) for the AGN component.
\item Col (5):    Scale length $r_s$ (kpc) of the exponential extended disk.
\item Col (6):    Parametric magnitude (mag) of the extended disk in the {\it I}-band.
\item Col (7):    S{\'e}rsic index for the bulge component in the best GALFIT model. Multiple values 
means that GALFIT cannot distinguish between them.
\item Col (8):    Effective radius of the bulge component in the S{\'e}rsic model.
\item Col (9):    Parametric magnitude (mag) of the bulge in the {\it I}-band.
\item Col (10):  Bulge-to-total host galaxy luminosity ratio.
\item Col (11):  Other components in best GALFIT model besides the extended disk and bulge. Details are given in Table \ref{bardisk}.
\end{tablenotes}
\end{threeparttable}
\end{sidewaystable}

\subsection{Determining the sky level}

\label{sec:sky} The sizes of the images are $800\times800$ pixels and
most galaxies extend to a radius of $\sim 100-200$ pixels or $\sim
5\arcsec-10\arcsec$.  To determine the sky level we start with a
circular region with a radius of $350$ pixels located at the center of
the galaxies.  In this way, on the one hand, we can exclude the noise
near the edges of the images, while on the other hand we can make use
of a large number of pixels.  Then we use the command {\it ellipse}
within IRAF, which follows the methods described in
\cite{Jedrzejewski1987}, to derive the one-dimensional radial profile
of the galaxy.  Obvious contaminating features such as bright stars
and small galaxies are masked.  Typically, the radial profile will
converge to an almost constant value at the outer radii. We take the
average value of the $2-4$ outermost radial bins in the profile to be
the sky value, which will be used in our fitting.  We also measure the
fluctuations in values at each radius in these outermost bins to
estimate the random error, which is typically $\sim 1\%$ of the sky
value. When there are contaminants nearby, the standard deviation can
reach 5\% of the sky value.

In cases where the primary galaxy is compact and there are multiple
stars or galaxies in the field, we use a smaller region for the sky
determination.  In cases where the galaxy is very extended (e.g., SDSS
J110501.97+594103.6) we use observations in the three WF chips to
determine the sky value.  Finally, there are some galaxies with
inclined disks that almost fill part of the image while leaving other
parts almost empty.  In these cases we put rectangular regions along
the galaxy minor axis and use these regions to measure the sky counts
and standard deviation.

As the measured galaxy sizes are very sensitive to the adopted sky
value, we check the sky values in a few ways. For the 17 most extended
galaxies, we create a radial profile extending beyond the PC chip, and
calculate the average sky value where the profile becomes constant.
In addition, we calculate the average counts in a few randomly chosen
blank regions. Results from these different methods agree with the
adopted values within two sigma, which means the sky values and
uncertainties we adopt are well-defined. There is another way to
measure the sky value using the mode, or most common pixel value.
This determination should be less affected by contamination.  The sky
values determined from the mode agree with our adopted value within
one standard deviation.  Another consistency check is described in the Appendix.

\subsection{Parametric fitting with GALFIT}

Following \cite{Greenetal2008}, we use the GALFIT routine to perform
full two-dimensional profile decompositions (e.g.,
\citealt{Pengetal2002}; \citealt{Pengetal2010}) with the following 
goals.  First, we want to
determine what fraction of the galaxies contain various components,
such as bulges, disks, and bars.  Second, we wish to measure the bulge
properties (luminosity and size) to study the bulge scaling relations
(e.g., the fundamental plane) for these galaxies.  Third, we will
quantify the morphology of the galaxies based primarily on the
bulge-to-total (B/T) light ratio (e.g.,
\citealt{SimiendeVaucouleurs1986}).  Two-dimensional fitting has the
benefit that different components can be modeled with different
position angles.  PSF convolution is included in a straightforward
way.  Also, complex components such as nuclear disks and bars can be
included.  GALFIT models the galaxy components as axisymmetric
ellipsoids.  Lopsided components can also be modeled with the most
recent version, GALFIT 3.0 (\citealt{Pengetal2010}), which we use
here\footnote{More information about GALFIT can be found at GALFIT
  home page online:
  http://users.obs.carnegiescience.edu/peng/work/galfit/galfit.html.}.

In general, we proceed as follows. First, we look at the image and the one
dimensional radial profile to determine what components to include in
our model.  Sometimes there is an obvious bar or disk in the galaxy,
which is included in the fitting. Most of the time, there are no
distinguishing structures and we perform an initial
fit with a central PSF for the AGN and a generalized S{\'e}rsic model
\begin{equation}
\Sigma(r)=\Sigma_e\
\text{exp}\left\{-b_n\left[\left(\frac{r}{r_e}\right)^{1/n}-1\right]\right\},
\label{sersic}
\end{equation}
where $r_e$ is the effective (half-light) radius, $\Sigma_e$ is the
surface brightness at $r_e$, $n$ is the S{\'e}rsic index, and $b_n$
is chosen so that the region within $r_e$ contains half of the light
in the profile integrated to infinity.

For $n=1$, the S{\'e}rsic model is reduced to an exponential
profile. Exponential profiles are usually used to describe disks.
The exponential profile is written in another form:
\begin{equation}
\Sigma(r)=\Sigma_0\text{exp}\left(\frac{r}{r_s}\right),
\end{equation}
where $\Sigma_0$ is the central surface brightness and $r_s$ is the
radius at which the surface brightness drops to $1/e$ of the central
surface brightness. In this paper, the disk size is reported with
$r_s$ while for other components, we report $r_e$. For $n=1$, there
is a simple relation $r_e=1.678r_s$.

For $n=4$, the S{\'e}rsic profile is equivalent to a
\cite{Vaucouleurs1948} profile, which is often used to describe the
light profile of elliptical galaxies. In GALFIT, if we let the
S{\'e}rsic index $n$ float freely, GALFIT will sometimes adopt an
unacceptably large $n$ to reduce the $\chi^2$. As shown in
\cite{BlantonMoustakas2009}, most nearby galaxies have a S{\'e}rsic
index smaller than 5. During our fitting, we fix $n = 1,2,3,4$
respectively for the bulge component and see which gives the best
fit. Other groups \citep[e.g.,][]{Bensonetal2007,SimmonsUrry2008} adopt 
a similar strategy of fixing the S\'ersic index to discrete values.
If different $n$ values give almost the same parameters and
the same reduced $\chi^2$, then we cannot distinguish between the different
S{\'e}rsic models. For the bar component, we model the
intensity distribution with $n=0.5$ (e.g., \citealt{Freeman1966};
\citealt{Greenetal2008}), which is a Gaussian profile.

\begin{figure}
\hspace{-2.0cm}
\includegraphics[width=1.2\hsize]{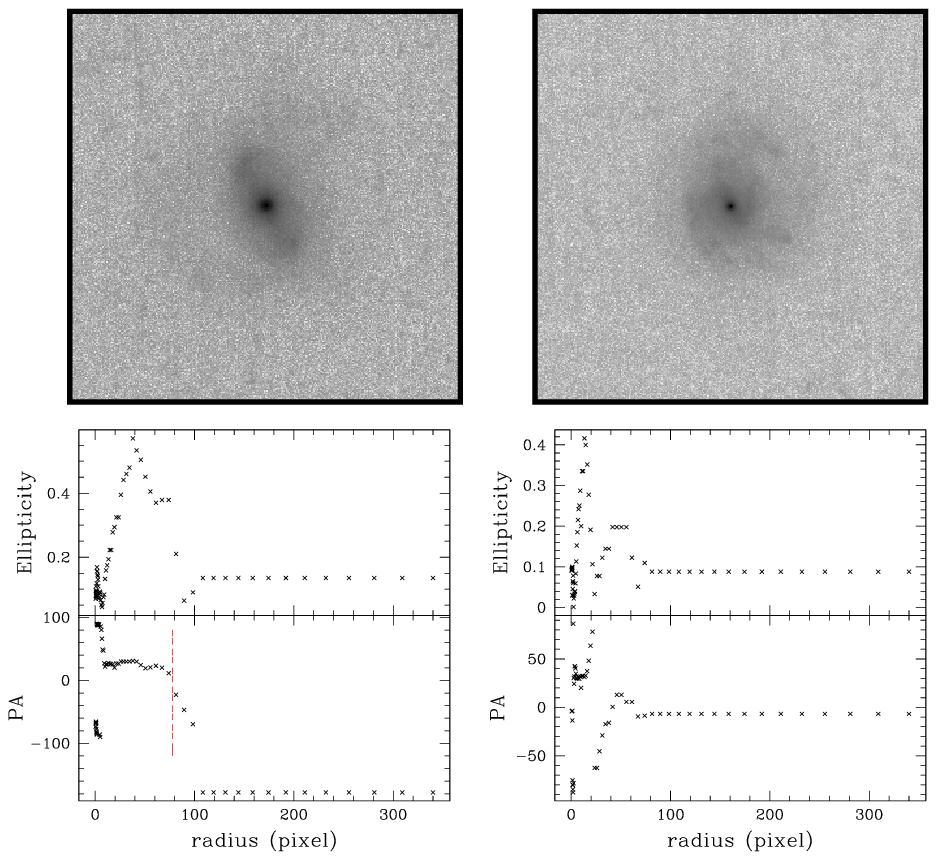}
\vspace{-12cm}
\caption[]{Example ellipticity and PA profiles
that are used to determine whether or not a given galaxy contains a bar. 
The left panels are for 
galaxy $0823 +0606$ while the right panels are for galaxy 
$0833 + 0620$. The top two images are $300\times300$ pixels in size.
In the bottom-left panel, the sharp drop in ellipticity around $60$ pixels 
(red dashed line) and the constant 
PA confirm the bar we see in the image. In the bottom-right panel, 
although we see changing ellipticity and PA around 
$50$ pixels, we do not see any bar-like structure in the image. The 
change in ellipticity is likely influenced by the spiral arms so that the PA 
is not constant. } \label{bar}
\end{figure}

In order to determine whether a galaxy has bar structure, we visually
check the images one by one, since a weak bar may exist even when the
outer parts of the galaxy are well fit by a single $n=1$ component. For
the ambiguous cases when we cannot distinguish between a bar or
edge-on disk visually, we check the ellipticity profile and position-angle 
profile as described in \cite{Menendezetal2007}. We use the {\it
  ellipse} command in IRAF to plot ellipticity and position angle at
each radius.  For the galaxies with a round bulge at the center, we
should see a monotonic increase in ellipticity and a constant position
angle (PA) across the bar. Because the bar and the disk usually have
different PAs, we should also see the ellipticity drop abruptly while
the PA changes sharply at the end of bar.  If we see those signatures
simultaneously, we conclude that there is a bar.  If either signature
is missing and we do not see a bar clearly by eye, we do not classify
the galaxy as barred.  In Figure \ref{bar}, we show two examples of this
procedure. For galaxy $0823+0606$, we see both a sharp drop of the
ellipticity profile and an almost constant PA profile indicative of a
bar.  For galaxy $0833+0620$, although there is a change in the
ellipticity profile at $\sim 50$ pixels, the PA changes with radius and
we also do not see a bar-like structure in the image. This galaxy
has no large-scale bar.

After we find the best fit with the AGN component and one S{\'e}rsic
model, we compare the radial profile of the model and the image and
decide whether we need to add a new component. We continue in this way
until the model fits the image.  Our criteria are that there be no
large features in the one-dimensional residuals and that the reduced
$\chi^2$ has a minimum of $\sim$ one. 

Note that we are not fitting the spiral arms or knots of star
formation and other nonaxisymmetric features, which are apparent in
some of the residual images. However we do fit additional compact
components in the central region if necessary, which means any
component with an effective radius smaller than the $r_e$ of bulge. In
some cases, without this nuclear component there is also extra light
in the residuals regardless of the value of n.  A similar nuclear
component with size $\sim 100$ pc is also found in the nearby low-mass
AGN host POX 52 (e.g., \citealt{Barthetal2004,Thorntonetal2008}).
This is different from nuclear star clusters in late-type galaxies, 
which have typical radii $\sim2-5$ pc \citep[e.g.,][]{Bokeretal2004}. 
As the median redshift of our sample is $0.085$, a nuclear 
star cluster will fall within a single WFPC2 pixel. 
The bright nuclear point source only makes the situation worse. 
So  unfortunately, 
we do not have sufficient angular resolution to search for nuclear star clusters in 
our sample.

\section{Uncertainties and Upper Limits}
\label{sec:uncertainty}
In our GALFIT modeling there are two factors that dominate the
fitting uncertainties: the PSF model and the sky level.  Uncertainty
in the magnitude of the AGN component is mainly due to uncertainties
in the PSF model, while the sky value gives the biggest uncertainties
in the sizes of extended components. When running GALFIT, we use the
PSF model generated by Tiny Tim for each galaxy. To estimate the
uncertainty due to the PSF model, we fit all the galaxies again with
the other two PSF models as shown in Figure \ref{comparePSF}.  The
alternative PSF models have a similar radial profile to the Tiny Tim
PSF with some differences in the wings. When we refit the galaxies
with alternate PSF models, all fitting parameters are fixed except the
PSF model and the component magnitudes and sizes.  Then we calculate
the differences between the magnitudes and sizes of different
components between the new fits and the original fits, which are
caused by different PSF models.

To estimate uncertainties due to the sky level, we change the
sky value and fix all other parameters to their best-fit values 
(keeping the same PSF model). We test three different sky values.
First,  we allow the sky to be a free parameter in
GALFIT. Then we increase and decrease the sky level by one standard deviation 
(see \S \ref{sec:sky} for details of the sky determination).
Then we calculate the differences in the magnitudes and sizes of
different components between the new fits and the original fits.
The largest difference is taken to be the measurement
uncertainty, which is given in table \ref{parameters} and the
original fitting results are reported as our best fits.

For galaxies with no detected bulge or conversely no extended disk component, we
estimate an upper limit on these components. The effective size of the
undetected component is determined according to the relationship
between bulge and disk size given in \cite{MacArthuretal2003}, which
is $\langle r_e/r_s\rangle=0.22\pm0.09$. Here $r_e$ is the effective
size of the bulge component while $r_s$ is the size of the extended
disk. Take, for example, a galaxy with a detected bulge component but
no detected extended disk.  In GALFIT, we add an exponential disk
component with the size fixed according to the above relation and the
measured bulge size.  The axis ratio of the disk component is fixed to
unity as we want to estimate the upper limit for a face-on disk. The
parameters of all the other components in GALFIT are kept fixed. In
each case, we fix the magnitude of the disk component to be a certain
value and we let GALFIT calculate the new $\chi^2$.  We increase the
magnitude of the undetected component until $\chi^2$=1 (or until
$\chi^2$ increases by 10\% in cases where $\chi^2$ starts out larger
than 1).  This magnitude is taken to be the upper limit of the
undetected disk.  The upper limit on an undetected bulge component is
determined in a similar fashion. 
There are also six galaxies that do not include an AGN component 
for the best fits. We estimate an upper limit on the AGN in a similar fashion, 
by increasing the PSF magnitude until the final $\chi^2=1$. This PSF magnitude 
is taken as the upper limit of an AGN component. Two of the six 
galaxies are labeled as `c'  by \cite{Greenetal2007}, which 
means the broad line masses are very uncertain. All six  
galaxies have very bright bulge or disk components, which 
makes the AGN magnitude very Êuncertain.
The chosen value of $\chi^2$ that we
use to estimate the upper limit is a bit arbitrary. Nevertheless, this
procedure gives us some useful information on the magnitude of the
undetected component.

In order to minimize the effect of systematic uncertainties, we have
followed several general principles during the fitting.  We have
looked at all the images one by one to identify the bar and spiral
structures.  The disk component is always fitted by an exponential
profile as in nearby inactive galaxies, although disk profiles do vary
at large radius \citep[e.g.,][]{Pohlenetal2004,Erwinetal2005}.  In
this way we will not be confused by different choices of S{\'e}rsic
index for the disk.  Whenever we decide that there is a bar component, we
also add a disk component to get a physically reasonable model. During
the fitting, we guess the sizes of each component based on the 1D
profile and take the guess as an initial value for GALFIT so that it
will converge on reasonable results.  For additional checks on the 
photometry, see Appendix A.

\begin{center}
\begin{longtable}{ c c  c   }
\caption{Galaxies with bar structure}
\label{bardisk}\\
\hline \hline

  &\multicolumn{2}{c}{Bar}  \\
  \cline{2-3}\\
 
  & $m_I$  & $r_e$  \\
Name  & (mag) & (kpc)  \\
 (1)   &  (2)  &  (3)     \\
 \hline
 \endfirsthead

\multicolumn{3}{r}{{Continued on next page}} \\ \hline
\multicolumn{3}{l}{{Col (1): Abridged SDSS name.}}\\
\multicolumn{3}{l}{{Col (2): $I$-band magnitude of the bar component.}}\\
\multicolumn{3}{l}{{Col (3): Effective size of the bar component.}}\\
\hline

\endfoot
\hline
\endlastfoot

$0304+0028$ &   16.22$\pm$0.22 &  2.64$\pm$0.12  \\  

$0731+3926$  &  17.93$\pm$0.04 &  4.62$\pm$0.04  \\   

$0748+4540$  &   19.13$\pm$0.05 & 4.88$\pm$0.05 \\   

$0750+3157$ &  17.92$\pm$0.04 &  4.56$\pm$0.05 \\  

$0806+2419$ &  18.08$\pm$0.32 &  0.48$\pm$0.02 \\  

$0815+2506$ &  19.73$\pm$0.09 &  1.04$\pm$0.04 \\  

$0818+4729$ &  18.72$\pm$0.02 &  0.85$\pm$0.01 \\  

$0823+0651$  &   19.01$\pm$0.12 &  4.73$\pm$0.19 \\  

$0823+0606$  &  18.45$\pm$0.16  & 3.70$\pm$0.06   \\  

$0824+0725$  &  16.90$\pm$0.03  & 6.04$\pm$0.67  \\    

$0830+0847$  &   19.47$\pm$0.07 &  2.25$\pm$0.06 \\  

$0843+3610$ &   16.70$\pm$0.16 &  10.63$\pm$0.27 \\  

$0847+3604$ &   18.56$\pm$0.12 &  2.56$\pm$0.03 \\  

$0854+0808$ &   18.38$\pm$0.10 &  8.05$\pm$0.55 \\    

$0900+4327$ &   17.22$\pm$0.05 &  4.16$\pm$0.37 \\   

$0903+4639$ &  19.18$\pm$0.07 & 1.96$\pm$0.05 \\   

$0910+0408$ &   18.96$\pm$0.14 &  4.23$\pm$0.14   \\    

$0925+0502$  &  19.10$\pm$0.11 &  2.72$\pm$0.09   \\  

$0927+0843$  & 18.21$\pm$0.33   & 2.67$\pm$0.15  \\   

$0933+5347$ &  18.22$\pm$0.14 &  4.31$\pm$0.16 \\  

$0940+0324$ &  17.62$\pm$0.05 &  1.97$\pm$0.02 \\    

$0942+4800$ &  19.69$\pm$0.04 &  2.07$\pm$0.06 \\   

$0942+0838$ &   18.17$\pm$0.19 &  7.83$\pm$0.25 \\  

$0953+3650$ &   19.90$\pm$0.17 &  0.47$\pm$0.06  \\  

$0953+5626$ &   18.39$\pm$0.02 &  1.74$\pm$0.01  \\   

$1022+3837$ &   19.76$\pm$0.07 &  3.50$\pm$0.07 \\    

$1029+4314$ &   17.80$\pm$0.05 &  4.13$\pm$0.08  \\  
$1043+5121$  &  16.95$\pm$0.03 &  5.56$\pm$0.07  \\  

$1048+4133$ &  18.32$\pm$0.07 &  3.71$\pm0.04$  \\ 

$1051+6059$ &  17.68$\pm$0.09 & 6.00$\pm0.13$ \\  

$1057+4825$ &  18.06$\pm$0.03 & 1.78$\pm0.01$  \\  
 
 $1102+4638$  &   18.14$\pm$0.14 &  0.53$\pm$0.02 \\ 
 
 $1105+5941$  &   15.52$\pm$0.02 &  1.74$\pm$0.01   \\  
 
 $1123+4331$  &   17.87$\pm$0.07 &  3.49$\pm$0.12     \\  
 
$1126+5134$ &   16.56$\pm$0.07 &  1.96$\pm$0.03     \\  
 
   $1151+5613$ &   17.39$\pm$0.12 &  3.59$\pm$0.10     \\  
   
   $1153+4612$  &   16.25$\pm$0.07 &  3.56$\pm$0.04   \\  
   
   $1258+5225$  &   19.94$\pm$1.31 &  5.37$\pm$0.37     \\  
    
    $1313+0519$  &  17.73$\pm$0.43 &  2.89$\pm$1.23    \\  
    
  $1325+5429$  &  19.12$\pm$0.16 &  3.77$\pm$0.27    \\  
     
 $1342+4827$  &  19.58$\pm$0.06 &  1.39$\pm$0.08     \\   
     
  $1416+5528$  &   16.90$\pm$0.30 &  6.65$\pm$0.18    \\  
      
$1435+3413$  &   18.71$\pm$0.04 &  1.22$\pm$0.01     \\  
       
$1437+5458$  &   18.22$\pm$0.02 &  0.77$\pm$0.04   \\ 
   
$1506+3413$   &   17.23$\pm$0.04 &  5.30$\pm$0.19   \\  

$1546+4751$   &   18.97$\pm$0.04 &  2.90$\pm$0.01    \\  
 
$1617-0019$   &   19.83$\pm$0.05 &  1.00$\pm$0.02     \\  

$1621+3436$   &   18.55$\pm$0.07 &  3.58$\pm$0.08   \\  
 
$1708+6015$   &   18.22$\pm$0.16 &  9.34$\pm$0.09      \\  

$2137-0838$   &   19.38$\pm$0.10 &  3.32$\pm$0.30    \\  

$2211-0105$  &   19.18$\pm$0.05 &  5.45$\pm$0.23    \\  
 
$2238+1433$  &   18.80$\pm$0.06 &  4.80$\pm$0.03     \\  

$2358+0020$ & 19.56$\pm$0.07  & 4.56$\pm$0.06     \\  
\hline \hline
\end{longtable}
\end{center}


\begin{center}
\begin{longtable}{ c c  c  }
\caption{Galaxies with Compact Nuclear Components}
\label{nuclear}\\
\hline \hline
 &\multicolumn{2}{c}{Compact Nuclear Component}  \\
  \cline{2-3}\\
 & $m_I$ (mag) & $r_e$  \\
Name  &  & (kpc)   \\
 (1)   &  (2)  &  (3) \\
\hline
\endfirsthead

\multicolumn{3}{l}{{Col (1): Abridged SDSS name.}}\\
\multicolumn{3}{l}{{Col (2): $I$-band magnitude of the nuclear component.}}\\
\multicolumn{3}{l}{{Col (3): Effective size of the nuclear component.}}\\
\hline

\hline
\endlastfoot

$0824+2959$ &   16.63$\pm$0.01 &  0.093$\pm$0.001 \\   

$1153+5256$  &  18.62$\pm$0.22 &  0.91$\pm$0.22  \\  

$1223+5814$  &  17.54$\pm$0.02 &  0.148$\pm$0.004     \\  

$1656+3714$ &  19.63$\pm$0.40 &  0.17$\pm0.06$  \\  

\hline
\end{longtable}
\end{center}


\begin{sidewaysfigure}
\centering
\epsfig{file=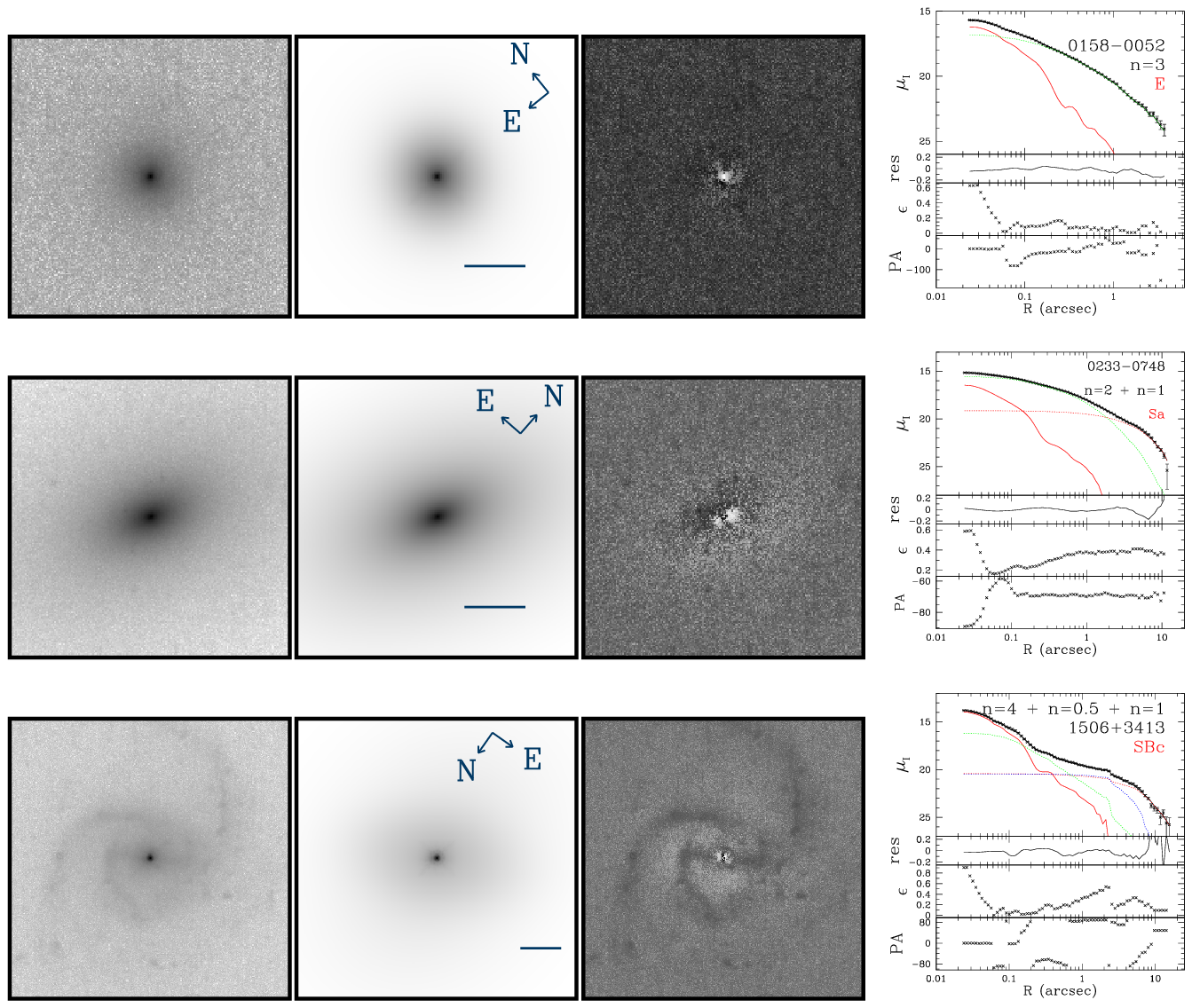, angle=-90, width=1.1\hsize}
\vspace{-0.8cm}
\caption{Examples of \emph{HST} images and GALFIT decompositions. From left to
right, we show the \emph{HST} image, the model image, the residual image and the 
one dimensional profile.  We plot the radial profile of the AGN (solid red line), the bulge 
(dashed green line), the extended disk (dashed red line), and the bar (dashed blue line).
The S{\'e}rsic index of each component is labeled 
in order of increasing size ($r_e$) of that component. These six
examples include galaxies with most combinations of different components. 
Note that the stretch of the residual images is different from the original images. 
The counts in the residual images are only $\sim\pm2\%$ of those in the original images.
Similar figures for all objects are included in the electronic edition of this paper.
} \label{exampleimage}
\end{sidewaysfigure}

\begin{sidewaysfigure}
\ContinuedFloat
\epsfig{file=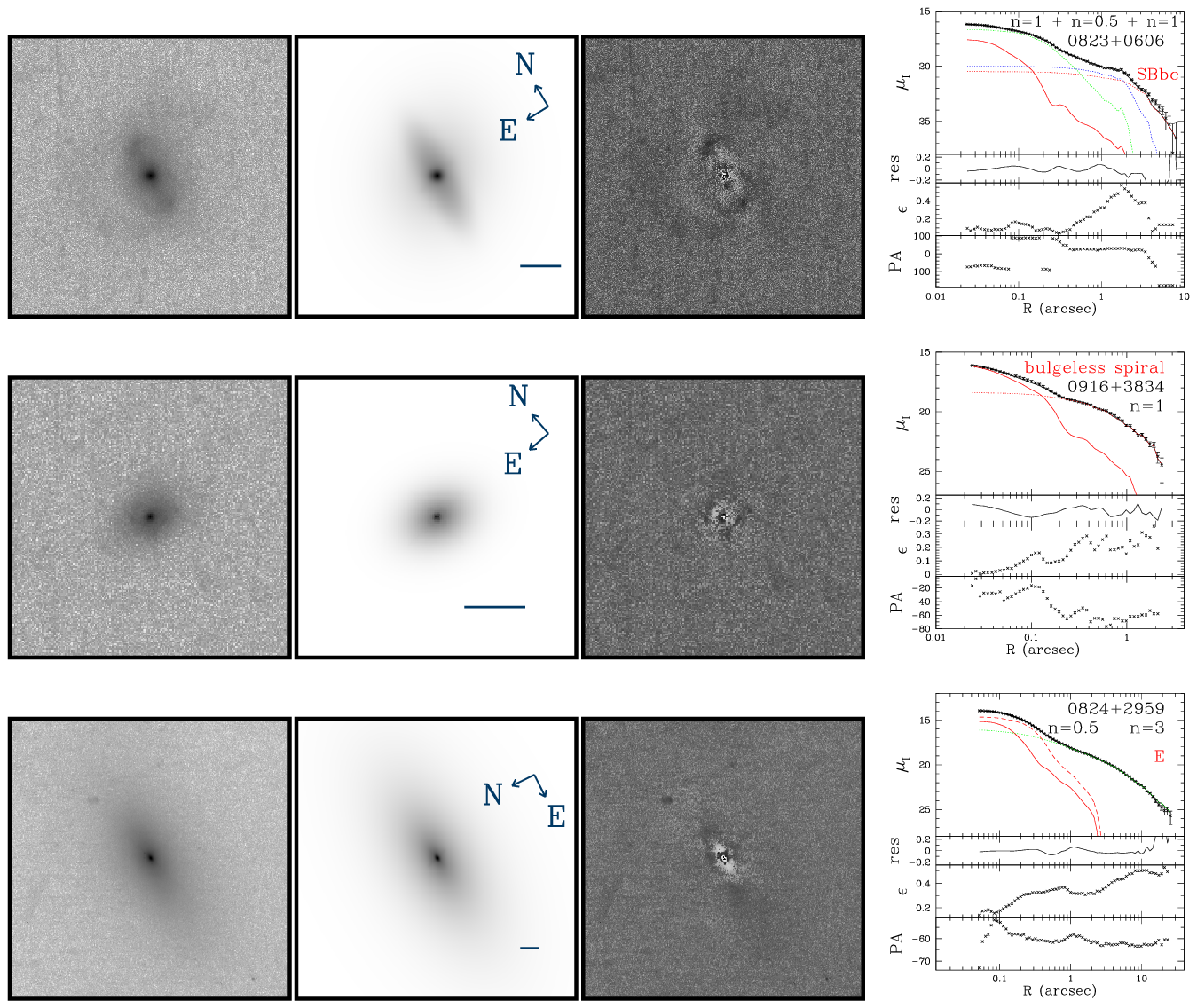, angle=-90, width=1.15\hsize}
\caption[]{Continued.}
\end{sidewaysfigure}

\begin{figure}
\centering
\includegraphics[width=1.0\hsize]{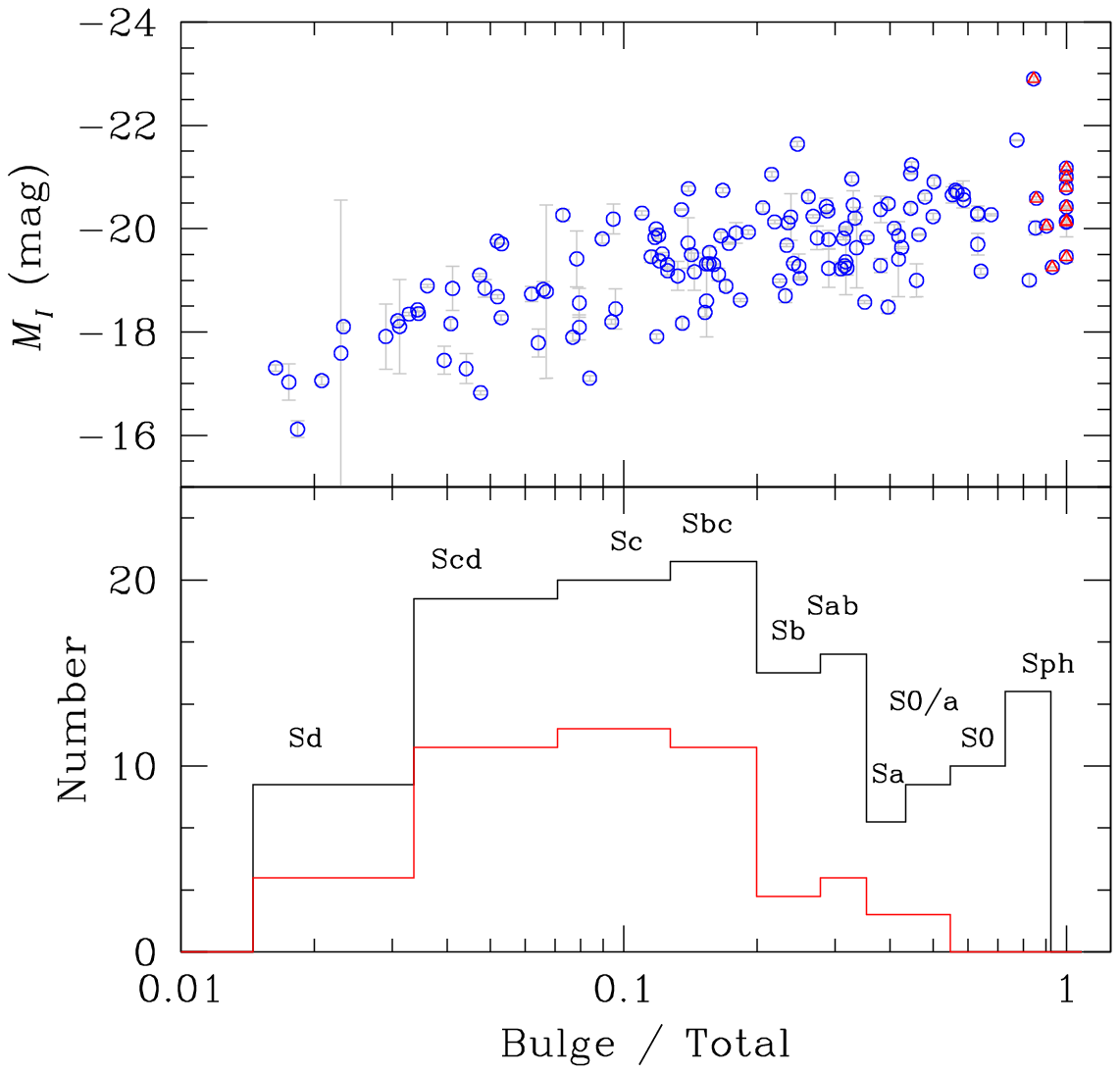}
\vspace{3mm} \caption{{\it Top}: Relation between 
bulge-to-total ratio and bulge magnitude. 
Galaxies without a detected extended disk
are labeled with red triangles (note that those with a bulge-to-total 
ratio less than one have a nuclear component).
{\it Bottom}: Histogram of the bulge-to-total ratio, from 
which we can see the distribution of galaxies in terms of Hubble 
type. The red line is the histogram for galaxies with bars. Note that the 
galaxies labeled with ``Sph" have Hubble type ``E". But they 
are not elliptical galaxies. Instead, they are similar to spheroidal galaxies 
according to their positions in the fundamental plane. }
 \label{BTratio}
\end{figure}

\section{Galaxy Morphology}
\label{sec:morphology}

We have selected a sample of galaxies with the lowest BH masses known.
Based on the SDSS imaging, the galaxies have relatively low masses as
well, with magnitudes $\sim 1$ mag below $L^*$.  Now, with detailed
image decompositions in hand, we are in a position to address the
morphology of the host galaxies.  Galaxy structure (e.g.,
S\'ersic index, presence of bars or nuclear spirals) will help us
determine whether each bulge is a classical or pseudobulge
\citep{Kormendyetal2004,FisherDrory2010}.  We quantify the bar
fraction, since bars have been suggested as a feeding mechanism for
low-level AGN activity \citep[e.g.,][]{Shlosmanetal1990}. We also
count the number of galaxies with interacting companions.

\subsection{Statistical Results}
\label{sec:statistic}
With our image decompositions we only fit the main components
of the galaxies, including bulges, extended disks, bars and nuclear
structures.  Spiral and irregular structures such as nuclear spirals or 
rings are not fitted and those structures can be seen in
the residual images. In Figure \ref{exampleimage}, we show some example
\emph{HST} images, including the model image, the residual image
and a radial profile\footnote{\emph{HST} images for the full list can
  be found in the online version of the paper.}.  The models generally match the
data well.

In our sample of 147 galaxies, 136 ($93\%$) have extended disks. Of
the disk galaxies, $53$ of them ($39\%$) have bars.  The bar
properties are shown in Table \ref{bardisk}.  Of the galaxies with
disks, only seven are consistent with having no bulge component at
all. The remaining 11 diskless systems are smooth and featureless
galaxies.  Based on their position in the fundamental plane (\S
\ref{sec:FPrelation}), we will argue that these galaxies are
spheroidal, rather than elliptical, galaxies.  Half of the diskless
galaxies have S{\'e}rsic index $n>2$, which is larger than the average
S{\'e}rsic index of the whole sample but similar to POX 52
\citep[e.g.,][]{Thorntonetal2008}.  However, in general their
luminosities and sizes are larger than POX 52.  The average absolute
bulge luminosity and bulge size of these $11$ galaxies are both larger
than the average values of the whole sample. As listed in Table
\ref{nuclear}, four of those galaxies have an additional compact
component with a size that is smaller than the effective radius of the
bulge. Finally, there are seven galaxies consistent with having no bulge
component at all. Details are given below.

\subsection{Interacting galaxies}

\begin{figure}[htp]
\hspace{-2cm}
\includegraphics[width=1.2\hsize]{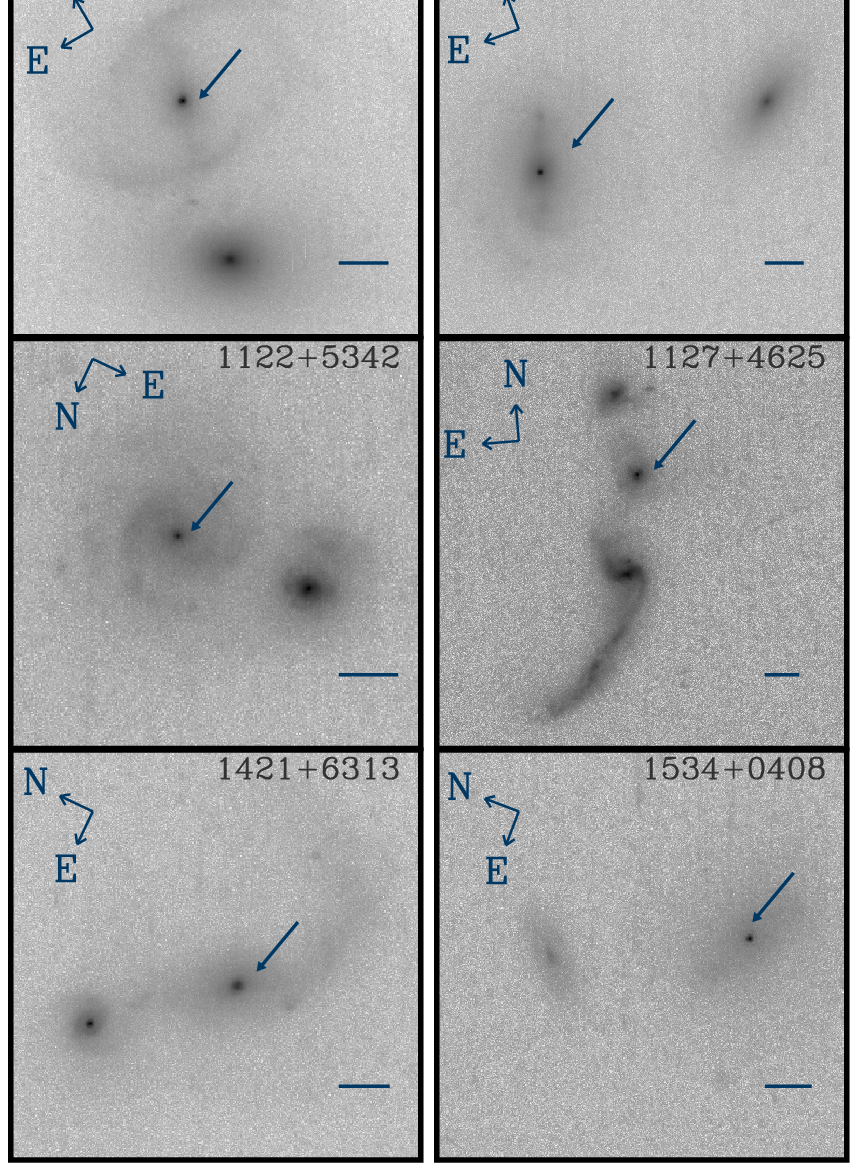}
\vspace{-8cm}
\caption[]{Interacting galaxies. Here we show six examples of
close interacting galaxy pairs. Some of them have obvious tidal tails
($1127+4625$ and $1421+6313$). There are 13 such
galaxies in our total sample of 147 galaxies. The arrow indicates the program 
galaxy in each case.  The scale bars are $2\arcsec$.} 
\label{interacting}
\end{figure}

Mergers are believed to be an efficient way to trigger AGN activity
and feed gas to the central supermassive BH (e.g.,
\citealt{HopkinsQuataert2010,Mayeretal2010}). Mergers are also thought
to trigger star formation in galaxies (e.g.,
\citealt{Bartonetal2000,Bartonetal2007,BlantonMoustakas2009}).
Observations have found examples of mergers in AGNs with more
massive BHs. (e.g., \citealt{Liuetal2010}).  For our sample of
low-mass BHs, we can check whether or not mergers are a common and
important phenomenon.

In our sample, 13 of the galaxies ($9\%$) are detected with close
companions that are likely physically interacting.  Some of them show
obvious long tails, which are clear signatures of tidal
interaction. In Figure \ref{interacting}, we show six example
interacting galaxies. Although the fraction of interacting galaxies in
our sample is only $\sim 9\%$, it is already larger than the value
$1\%-2\%$ reported for luminous galaxies (e.g.,
\citealt{BlantonMoustakas2009}).  However, we caution
that we identify the companions only visually based on our \emph{HST}
images. We do not use well-defined criteria, nor do we have a
well-defined control sample to compare with (e.g.,
\citealt{WoodsGeller2007,Dargetal2010}) as this is not our main focus.

\subsection{Bulge-to-total ratio}

The ratio between the bulge and total light (B/T, the AGN is
excluded), varies from 0 (no bulge detected) to $1$ (for
spheroidal galaxies).  The mean value is
$\langle$B/T$\rangle=0.28$ with a median of $0.18$.  In most cases B/T
is either less than $0.5$ or greater than $0.9$, with only 15 galaxies
in between. If we restrict our sample to those galaxies with a
detected disk component, then the mean becomes
$\langle$B/T$\rangle=0.23$ with a median value of $0.16$. In this
case, the ratio is always $<0.85$.

Based on B/T, we can assign a Hubble type to each galaxy following the
RC3 (Third Reference Catalogue; \citealt{SimiendeVaucouleurs1986};
\citealt{deVaucouleursetal1991}).  For the $129$ galaxies with a
detected disk and bulge component we find: $2\%$ are Hubble type E;
$15\%$ are Hubble type S0;
$18\%$ are Hubble type Sa; $28\%$ are Hubble type Sb; $30\%$ are
Hubble type Sc; and the remaining $7\%$ are Hubble type Sd. A
histogram for the Hubble type distribution is given in the bottom
panel of Figure \ref{BTratio}, which shows that our sample peaks at
Hubble type Sbc to Scd.

In the top panel of Figure \ref{BTratio}, we also show the relationship
between the bulge magnitude and B/T.  This plot clearly shows that B/T
is larger when the bulge luminosity is larger.  In other words, when
the bulge becomes more luminous, the disk component becomes relatively
faint.  It is clear that some of the bulge magnitudes are very
uncertain, particularly at low values of B/T.  The uncertainty in the
bulge magnitude is dominated by uncertainties in the PSF model but the
sky level also plays a role.

There are $27$ galaxies ($18\%$ of the total sample) with B/T $<5\%$.
As the extension of these very faint galaxies, there are even some
galaxies with no detected bulge component, which are described below.

\subsection{Bulgeless galaxies}

As discussed in the introduction, we are interested in how many
bulgeless galaxies contain AGNs. In our sample, we find seven galaxies
(5\% of the sample) that are best fit by pure disks. They are
$0916+3834$, $0942+4800$, $0953+3650$, $1102+4638$, $1116+4236$,
$1437+5458$, and $1534+0408$. These galaxies are either fitted by a
pure exponential disk (three) or a bar plus an exponential disk
(four).  The bulge magnitude limits range from $<20.6$ to $<18.4$ mag.
However, because there is a bright AGN component at the center and
four of them are at redshift $z \approx 0.2$, a small bulge could
escape detection in the \emph{HST} images. Deeper and
higher-resolution observations are needed to confirm their bulgeless
nature.  Such a small number of bulgeless galaxies with AGN activity
even in such a large sample suggests that these systems are truly
rare, although the SDSS selection does introduce a bias against faint
galaxies and AGNs \citep{Greenetal2007a}.

\subsection{Bars and Nuclear Structures}
\label{sec:bar}

In our sample, there are 53 galaxies with detected bars and four
with compact nuclear components (i.e., components smaller than the
bulge effective radius that require an additional S\'ersic profile). 
Looking only at galaxies with extended
disks, the bar fraction is $39\%$. Excluding the disk galaxies with an
inclination angle larger than $60\degr$, where it is difficult to
detect a bar (e.g., \citealt{Haoetal2009}), the bar fraction drops to
$37\%$. It is interesting to compare our sample with narrow-line
Seyfert 1 (NLS1) galaxies.  Like our galaxies, NLS1s are thought to
have BHs with relatively modest BH masses ($\le 10^7\msun$) and high
Eddington ratios (e.g., \citealt{Poundsetal1995,Crenshawetal2003}).
\cite{Crenshawetal2003} finds that $>60\%$ of the NLS1s in their
sample have a bar.  The bar fraction in our sample is
much smaller than that of \cite{Crenshawetal2003}.  Actually, the bar
fraction here is also smaller than that of
luminous AGNs and star-forming galaxies. Based on the classification
and structure information for $\sim2000$ galaxies from the SDSS,
\cite{Haoetal2009} claims that the AGN optical bar fraction is 47\%
and that the optical bar fraction in star-forming galaxies is 50\%. In
inactive galaxies, the bar fraction is only $29\%$, which is close to
that in our sample.

A relation between bar fraction and B/T is also claimed. People find
that the bar fraction increases with decreasing B/T, meaning that the
bar fraction is larger in disk-dominated galaxies (e.g.,
\citealt{Marinovaetal2009}). This conclusion is also confirmed in our
sample. We divide our sample into three equal bins in B/T: 0 to 0.33,
0.33 to 0.66 and 0.66 to 1. The bar fraction (with highly inclined
``bars" excluded) in the three bins is $45\%, 11\%$ and $0\%$
respectively. Histograms of the barred galaxy fractions for each
Hubble type (Figure \ref{BTratio}) shows that the bar fraction is
largest in Sbc-Scd galaxies and is very small for early-type galaxies.

Some theoretical models have proposed that bar driving can be an
efficient mechanism to funnel gas down to small scales where it can be
accreted by the BH
\citep[e.g.,][]{Shlosmanetal1990,HopkinsQuataert2010}. However,
observations do not find a definite connection between bars and AGN
activity. For example, some groups (e.g.,
\citealt{Hoetal1997,MulchaeyRegan1997}) do not find an excess of bars
in Seyfert galaxies, while other groups (e.g.,
\citealt{Knapenetal2000,Laurikainenetal2004}) claim a higher fraction
of bars in Seyferts. We have also compared the AGN magnitudes for the
galaxies with and without bars. The galaxies with bars do not show
significantly larger AGN luminosities.  We see no evidence that bars
are the dominant feeding mechanism for the AGNs in our sample. Instead, the
bar fraction is roughly consistent with what we expect from inactive
galaxies.

\section{Bulge Morphology}
\label{sec:pseudobulge}

We now turn to the galaxy centers, and we consider three types of
centrally concentrated components: classical bulges, pseudobulges, and
spheroidals.  For the galaxies without disks, we are trying to
distinguish between a small elliptical galaxy and a luminous
spheroidal.  We follow the definition of spheroidal galaxies from
\cite{Kormendyetal2009}.  Spheroidals are physically small, dynamically hot
stellar systems and thus are often called ``dwarf elliptical
galaxies".  While superficially they resemble elliptical galaxies,
they are much less dense at a given size or luminosity than low-mass
ellipticals.  Spheroidal galaxies and elliptical galaxies are located
along nearly perpendicular tracks in the fundamental plane
projections, presumably reflecting differences in formation history
(\citealt{Kormendyetal2009}).  In terms of structure, spheroidal
galaxies are closer to disk, rather than elliptical, galaxies 
\citep[e.g.,][]{kormendy1985}.  They
are thought to be defunct spiral galaxies that lost their gas through
interactions with a more massive companion.  We should note that the
interpretation we adopt here is not universally accepted in the
literature
\citep[e.g.,][]{Gavazzietal2005,Ferrareseetal2006,Coteetal2007}, but
we believe it is well justified by the observations
\citep{Kormendyetal2009}.

Turning now to the disk galaxies, it is widely accepted that there
are actually two kinds of bulges, which are formed in two different
ways (see the review in \citealt{Kormendyetal2004}).  Classical bulges
are thought to form in mergers. Such bulges are similar to scaled-down
elliptical galaxies in terms of their stellar populations and
structural properties.  

The other type of bulge is a pseudobulge.  Pseudobulges are believed
to be formed by secular processes, including the slow rearrangement of
material by bars, oval disks, and spiral structure. Pseudobulges
typically share several distinguishing properties (e.g.,
\citealt{Kormendyetal2004,Gadotti2009,FisherDrory2010}). First, they
usually have flatter shapes and have S{\'e}rsic indices $n<2$. They
also often have bars, spiral structures, or rings.  Second, they have a
high degree of rotational support, so that the ratio between the
rotation velocity and velocity dispersion is high \citep[e.g.,][]{KormendyIllingworth1983}. 
 As a corollary, they tend to have
smaller velocity dispersion $\sigma_{\ast}$ at fixed luminosity than
elliptical galaxies in the Faber-Jackson (1976) relation.  Third,
they have small B/T (e.g., \citealt{Kormendyetal2004}).  Fourth, they
are located at different positions in the $\langle\mu_e\rangle$ ---
$r_e$ plane, where $\langle\mu_e\rangle$ is the mean surface
brightness within the effective radius $r_e$.  In \cite{Gadotti2009},
the following relation is used to identify pseudobulges:
\begin{equation}
\langle\mu_e\rangle > 13.95+1.74\times\log (r_e/{\rm pc}),
\label{pseudorelation}
\end{equation}  
where $\langle\mu_e\rangle$ is measured in the SDSS $i-$band.
Finally, pseudobulges usually have young stellar populations and/or
recent star formation. Here, we use the B/T, S{\'e}rsic index,
and the presence of bars or rings to classify the bulges in our
sample. We warn the reader that there is not yet a clean, well-defined,
and widely accepted mathematical prescription to define pseudobulges,
especially with our limited data.
By combining these different criteria, we argue that most of the
bulges are likely to be pseudobulges.

Recall that our derived S{\'e}rsic indices are uncertain. However, in
$76\%$ of the galaxies, we can reliably distinguish S{\'e}rsic indices
larger or smaller than $2$.  In our sample, 33 galaxies are best fit
with $n>2$.  Excluding the seven galaxies without a bulge component, $79$
galaxies ($70\%$ of the galaxies with reliable S{\'e}rsic index) have
$n<2$ for the bulge.  Of the galaxies with a bulge component, $92\%$
have an extended disk and $\sim40\%$ have a bar
($\S\ref{sec:statistic}$, $\S\ref{sec:bar}$) . There are also
ring-like structures in $\sim17\%$ of our sample. As shown in Figure
\ref{BTratio}, $49\%$ of galaxies have B/T $<20\%$ and $74\%$ have B/T
$<40\%$. The remaining $14\%$ of galaxies have B/T $<5\%$.  Finally,
if we apply the empirical relation from Eq. \ref{pseudorelation}
\citep{Gadotti2009} to our sample, all of our galaxies with a bulge
component satisfy this relation. Thus, the majority of galaxies in our
sample have properties consistent with those of pseudobulges.

Now we are ready to examine where the galaxies lie in the Fundamental
Plane (e.g., \citealt{Kormendyetal2004}).  As the above evidence
shows, most of our galaxies are likely to be pseudobulges.  We
therefore expect them to scale differently in the fundamental plane
than elliptical galaxies \citep{Kormendy1980,
  Carollo1999,Kormendyetal2004,FisherDrory2010}.  Specifically, we
expect them to look more like disks.  In addition to the photometric
projections shown here, we also have stellar velocity dispersion
measurements and present the Faber \& Jackson (1976) relation in \S
\ref{sec:FJrelation}.

\begin{sidewaystable}
\begin{threeparttable}
\caption{Fit Results}
\label{fittingresult}
\begin{tabular}{ccccc|cccc}
\hline\hline
  & \multicolumn{4}{c|}{Our sample} & \multicolumn{4}{c}{Ellipticals and S0}  \\
\hline
Relation					& 	$\alpha$	& $\beta$	 &    $\epsilon_0$	&   median value & 	$\alpha$	& $\beta$	 &    $\epsilon_0$	&   median value \\
\hline
$\log(r_e)-\mu_e$				& 	$4.88\pm 029$ & $18.53\pm0.09$	&	0.16			&   $-0.26$	& 	$2.43\pm 0.16$ & $19.53\pm0.13$	&	0.67			&   0.097\\
$M_I-\mu_e$	&	$0.53\pm 0.12$ & $18.82\pm0.13$	&	0.54			&  $-19.42$	&	$-0.68\pm 0.10$ & $19.93\pm0.23$	&	1.31			&  $-20.97$\\
$M_I-r_e$			&	$-0.17\pm 0.02$ & $-0.23\pm0.02$	&	0.22			&  $-19.42$	&	$-0.33\pm 0.02$ & $0.24\pm0.05$	&	0.26			&  $-20.97$\\
\hline\hline
  & \multicolumn{4}{c|}{Our sample + G\"ultekin et al.} & \multicolumn{4}{c}{G\"ultekin et al.}  \\
\hline
Relation					& 	$\alpha$	& $\beta$	 &    $\epsilon_0$	&    median value	& 	$\alpha$	& $\beta$	 &    $\epsilon_0$	&    median value\\
\hline
$M_I-\log(\sigma)$			& 	$-0.13\pm 0.01$ & $2.00\pm0.02$	&	0.07			&   $-20.37$	& 	$-0.10\pm 0.01$ & $2.32\pm0.02$	&	0.06			&   $-22.52$\\
\hline
\end{tabular}
\begin{tablenotes}
\item Notes:  We fit $y = \alpha\times(x - x_0) + \beta$, where $x$ (the independent 
variable) is listed first in column 1 and $y$
(the dependent variable) is listed second.  $x_0$ is the median value of $x$, 
and is kept fixed.  $\epsilon_0$ is the intrinsic scatter. All the fits are done on the 
original data before corrections for stellar age are applied.
\end{tablenotes}
\end{threeparttable}
\end{sidewaystable}

 \subsection{The Fundamental Plane}
 \label{sec:FPrelation}

\begin{figure}[htp]
\centering
\includegraphics[width=1.0\hsize]{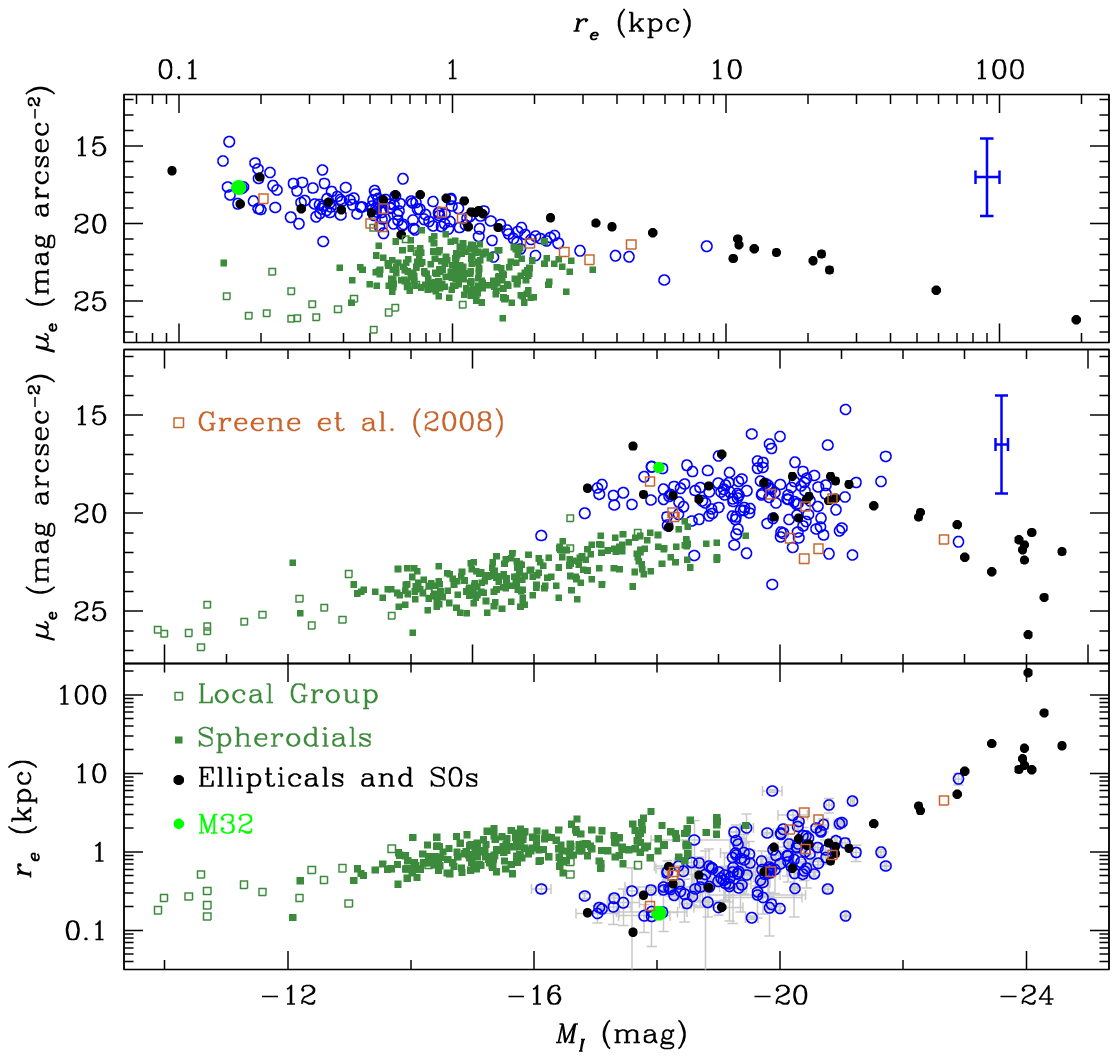}
\vspace{-0.4in} \caption{Fundamental plane projections of our sample, in blue 
open circles.
 We show the relations between the absolute magnitude of
the bulge $M_I$, the effective radius of the bulge $r_e$ and the
surface brightness at $r_e$, $\mu_{e}$. Bulge upper limits are not
shown.  For comparison, we show elliptical and S0 galaxies (filled
black circles) and M32 (large green circle) from
\cite{Kormendyetal2009}, Virgo spheroidal galaxies (green filled
squares) from \cite{Kormendyetal2009}, \cite{Ferrareseetal2006} and
\cite{Gavazzietal2005}, and Local Group dwarf spheroidal galaxies (open
squares) from \cite{Mateo1998} and \cite{McConnachieIrwin2006}. 
Their $V$-band magnitudes are shifted to the $I$-band
assuming $V-I=1.34$ mag (e.g., \citealt{Greenetal2008}). For
spheroidal galaxies, we use $V-I=1.09$ \citep{Fukugitaetal1995}. }
 \label{FPrelation}
\end{figure}

\begin{figure}[htp]
\centering
\includegraphics[width=1.0\hsize]{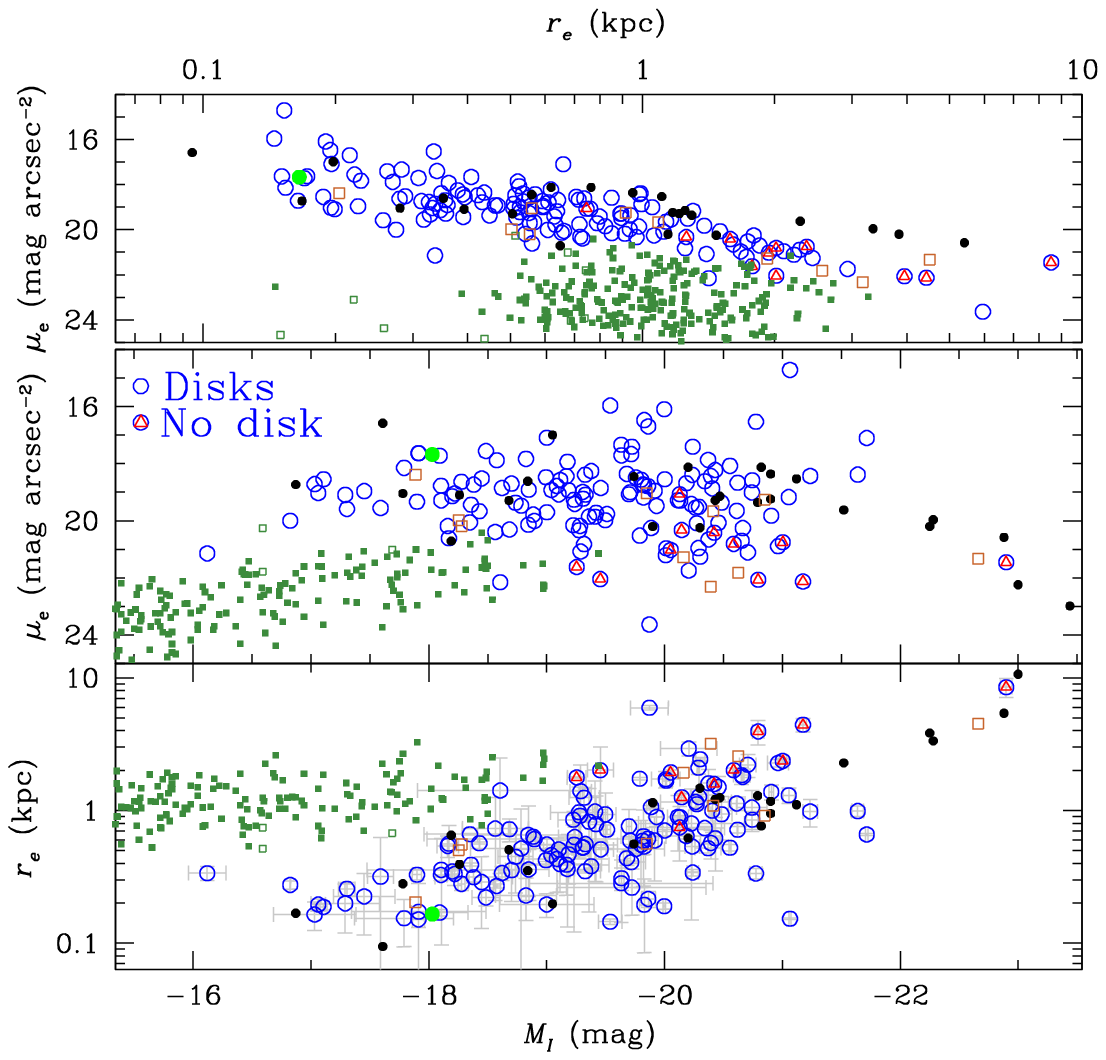}
\vspace{-0.4in} \caption{
An expanded view of the region containing our sample 
in Figure \ref{FPrelation}. All the symbols are the same as 
in Figure \ref{FPrelation}. We highlight galaxies without a detected extended 
disk (red triangles) and the $18$ objects from 
\citet[][brown squares]{Greenetal2008}.}
 \label{FPrelation2}
\end{figure}

\begin{figure}[htp]
\centering
\includegraphics[width=1.0\hsize]{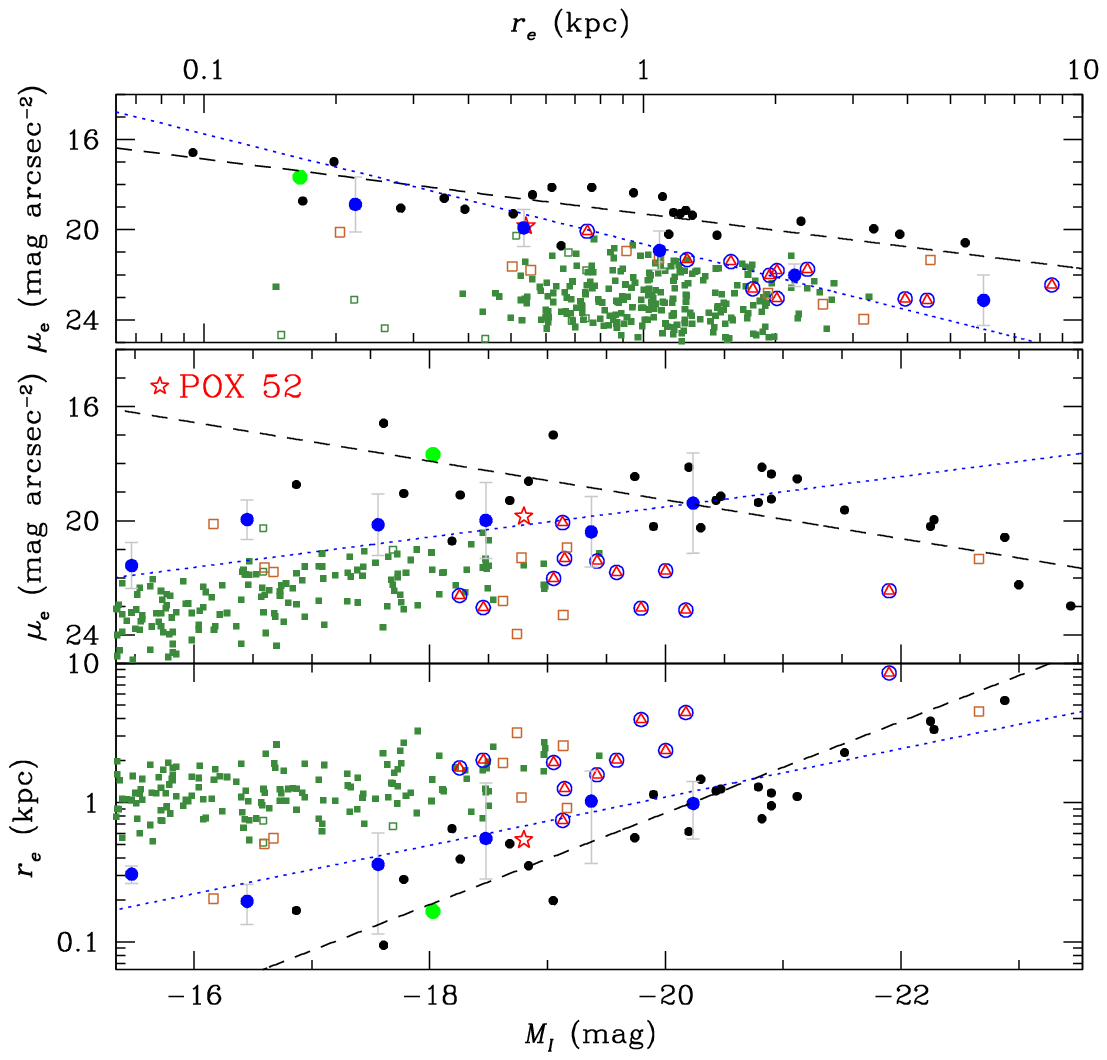}
\vspace{3mm} \caption{
The same fundamental plane plot as Figure \ref{FPrelation}, except that 
we shift the sample by $1$ magnitude to account 
crudely for differences in stellar populations between our sample, which 
likely contain some young stars, and the uniformly old elliptical galaxies. 
All other comparison samples are the same as in Figure \ref{FPrelation}. 
The filled blue circles are median values of $r_e$ and $\mu_e$ in uniform bins 
of $M_I$ or $r_e$ respectively for the galaxies with extended disks.  The 
dashed black lines are best fitting relations for elliptical galaxies and 
classical bulges (Table \ref{fittingresult}). The blue dotted lines are best 
fitting relations for our galaxies with extended disks with the $1$ mag shift 
applied (shifting the relations from Table \ref{fittingresult} by $1$ mag). 
}
 \label{FPrelation_shift}
\end{figure}

In Figure \ref{FPrelation}, we show three different projections of the
fundamental plane. The open blue circles are our sample.  All the
others are comparison samples taken in the $V$-band from
\cite{Kormendyetal2009}, \cite{Ferrareseetal2006} and
\cite{Gavazzietal2005}. We choose the Virgo galaxies as a comparison
here because they have very deep and uniform photometry and analysis. They
constitute a clean representative sample of local red galaxies.  The
sample is not complete, but nicely demarcates the region occupied by
elliptical and spheroidal galaxies in the fundamental plane.

In order to properly compare the samples, we calculate the magnitude
difference between the \emph{HST} filters F814W and F555W\footnote{The
  difference between the magnitude through the F555W filter and the
  $V$-band is smaller than $0.01$ mag, as shown in Table 3 of
  \cite{Fukugitaetal1995} and so we neglect the difference.} based on
a template elliptical galaxy spectrum from the Kinney-Calzetti atlas
\citep{Kinneyetal1996} using {\it calcphot} in IRAF.  The result is
$V-I=1.34$ mag, which is also the value used in \cite{Greenetal2008}.  We
shift the comparison elliptical and S0 galaxies from the $V$-band to
the $I$-band with this conversion factor. Spheroidal galaxies are
typically bluer than elliptical galaxies. For the spheroidal galaxies,
we adopt $V-I = 1.09$ mag, the color of Sbc galaxies listed in Table 3
of \cite{Fukugitaetal1995}, as spheroidal galaxies are redder than Scd
galaxies but bluer than Sa galaxies based on the $B-I$ color
\citep{Gavazzietal2005,Greenetal2008}.

We fit log-linear or linear relations to the fundamental plane for our
galaxies with disk components, following the same fitting method as
described in \cite{GreeneHo2006}. The fitting method is based on a
Levenberg-Marquardt algorithm.  The intrinsic scatter is an additional
error term that is chosen such that the minimum $\chi^2$ is one.
Upper limits are not included in the fits.  The fitting results are
shown in Table \ref{fittingresult}.  In the $\mu_e-r_e$ plane (top
left panel of Figure \ref{FPrelation}), our sample galaxies have a
steeper scaling than that of elliptical galaxies and classical bulges.
The difference is even more significant in the $\mu_e-M_I$ plane.  The
fundamental plane of these pseudobulges has a different slope and
normalization than the elliptical or the spheroidal galaxies, such
that the faint end is closer to faint elliptical galaxies while the
bright end is closer to bright spheroidal galaxies.

Previous results have also found that the photometric scaling
relations of pseudobulges diverge from those of classical bulges.
\cite{Carollo1999} find that their `exponential' (or pseudo) bulges
span a much wider range of effective surface brightnesses than do
classical bulges and elliptical galaxies. \cite{FisherDrory2010}
present a similar result.  They see almost no dependence of effective
surface brightness on effective radius.  Our findings are similar over
the range of effective radius that we share.  Thus, the observed
scaling relations also support our assertion that our disk sample is
dominated by pseudobulges.

There are 11 galaxies without extended disks.  These systems
are more challenging to interpret.  On the one hand, they are
typically 2 magnitudes more luminous than luminous spheroidal galaxies
(green squares in Figure \ref{FPrelation2}).  However, they are larger and have lower
surface brightnesses than faint ellipticals of their luminosity. 
The difference is most significant in $\mu_e-r_e$ plane (the top panel 
of Figure \ref{FPrelation2}).
Thus, as in \citet{Greenetal2008}, we conclude that the sample is
made up predominantly of spheroidal systems whose luminosities are
boosted by ongoing star formation.  As we will see in the next
section, their scaling in the Faber-Jackson relation strengthens our
conclusion here (Section \ref{sec:FJrelation}).

\subsubsection{Stellar population effects}

It is important to note that our sample galaxies likely contain
younger stars on average than Virgo cluster galaxies.  Unfortunately
we do not have direct measurements of the bulge colors.  However,
inactive spiral galaxies at these masses are all bluer than more
massive galaxies \citep[e.g.,][]{RobertsHaynes1994}. This is also 
true for bulges \citep[e.g.,][]{Benderetal1992}.
Furthermore, it is likely that some of the nuclear structures
represent ongoing or recent star formation.  One way to
compensate for this difference is to estimate the mass-to-light ratio
($M/L$) for the bulges.  We can estimate their dynamical $M/L$ using
$\sigma_{\ast}$ observations.  The bulge virial mass can be estimated as
$\sim 6.5\, r_e\sigma^2/G$ (e.g., \citealt{Taloretal2010}), where
$G$ is the gravitational constant. We find an average $M/L$ in our
sample of $\sim 1.8$ (in units of $M_{\odot}/L_{\odot}$). For a typical
elliptical galaxy in the $I$-band, the $M/L$ is $\sim 4-6$
\citep[e.g.,][]{Cappellarietal2006}.  If we keep the stellar mass
fixed but imagine that the stars evolve to the same
age as those in elliptical galaxies, the luminosity would
be decreased by a factor of $2.2-3.3$, corresponding to an
increase of $0.8-1.3$ mag.  Although we do not know the bulge
$M/L$ for individual galaxies, we can take the average
impact of fading the stellar populations to be $\sim$ one magnitude.

Applying a shift of one magnitude to our sample, the fundamental
plane would then appear as shown in Figure
\ref{FPrelation_shift}. This average accounting for stellar population
effects only strengthens our conclusions.  Our galaxies occupy a
different region of the fundamental plane from elliptical/S0 galaxies.
At the bright/large end, the galaxies clearly deviate from the
fundamental plane of massive elliptical galaxies.  However, at the
faint/compact end, the galaxies are more consistent with the
fundamental plane of small elliptical galaxies like M32 than they are
with spheroidals of the same size or luminosity.  In addition, the locus of 
the diskless galaxies moves closer to that of the spheroidal galaxies.

\subsection{The Faber-Jackson Relation}
\label{sec:FJrelation}

\begin{figure}[htp]
\centering
\includegraphics[width=1.0\hsize]{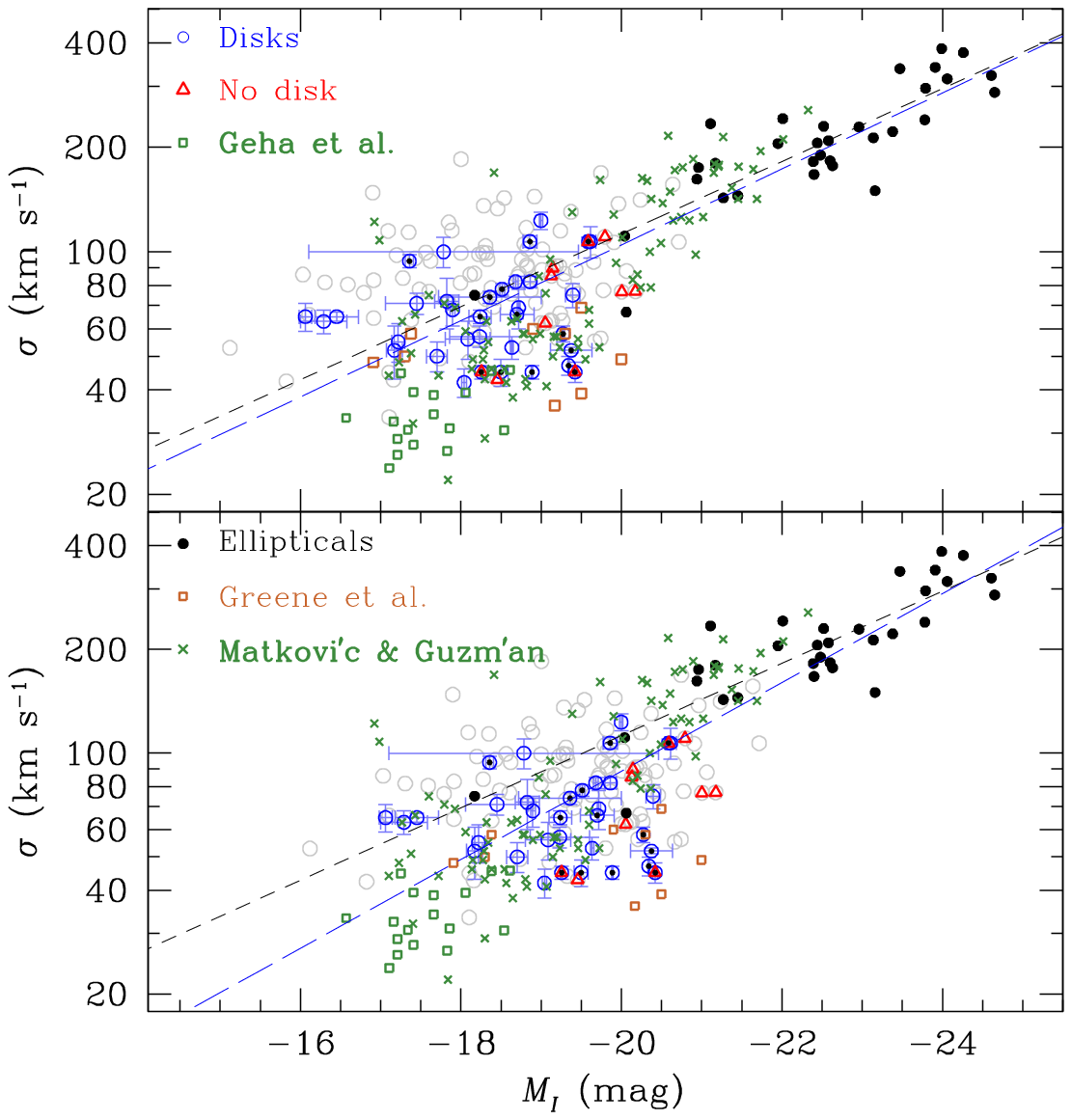}
\vspace{-0.2in} \caption{
{\it Bottom}: Faber-Jackson relation between bulge 
magnitude and velocity dispersion.  Open 
circles are our sample. Blue open circles have
stellar velocity dispersions from ESI/MagE measurements \citep{Xiaoetal2011} 
while grey open circles have velocity dispersions from [\ion{S}{2}] 
line widths in SDSS spectra \citep[][]{Greenetal2007}. 
We also highlight galaxies with no extended 
disk (red triangles) and those with low disk contamination (black solid 
circles; see \S 6.2.1 for details).  We show the fit to the 
comparison sample (black solid points) 
from \cite{Gutekinetal2009} as a short-dashed line and the best fit 
with our sample included as the long-dashed line. The open green 
squares and green crosses are faint early-type galaxies from \cite{Gehaetal2003} 
and \cite{MatkovGuzma2005} respectively.
{\it Top}: The bulge magnitudes for our sample are shifted by 
one magnitude to account for younger stellar populations.}
 \label{FJrelation}
\end{figure}

We have considered photometric projections of the fundamental plane.
For a subset of galaxies, we can also examine the Faber-Jackson (1976)
relation between stellar velocity dispersion ($\sigma_{\ast}$) and
bulge luminosity.  Previous results suggest that pseudobulges
demonstrate considerably larger scatter in the Faber-Jackson relation,
with a tendency to have a lower $\sigma_{\ast}$ at a given luminosity
\citep[e.g.,][]{KormendyIllingworth1983,DresslerSandage1983,
  Kormendyetal2004,Greenetal2008}. We examine the situation for
our sample.

We have $\sigma_{\ast}$ measurements for $34$ galaxies in the sample. 
The data were taken at Keck with ESI and Magellan with MagE.  The
measurements are based on direct-pixel fitting of broadened stellar
templates to the galaxy spectra \citep{Barthetal2005,Xiaoetal2011}.
For the remaining objects, we do not have direct stellar velocity dispersion measurements.
Instead, we use the widths of forbidden lines from the narrow-line
region as an approximate indicator of the stellar velocity
dispersion. In most AGNs, the widths of low-ionization emission lines
such as [\ion{O}{2}] and [\ion{S}{2}] are a good proxy for
$\sigma_{\ast}$, as shown in \cite{NelsonWhittle1996} and \cite{GreeneHo2005a}, 
and then discussed
in detail in \cite{Ho2009}.  In this paper, we refer to
dispersions based on [\ion{S}{2}] as $\sigma_{\text{gas}}$, which are
used whenever we have no $\sigma_{\ast}$ measurement.

\subsubsection{Restricted Sample}
\label{sec:cleansample}

In these disk-dominated galaxies, the observed $\sigma_{\ast}$ of the
bulge can be contaminated by rotationally dominated disk light.  At
Keck we employed a slit of $0\farcs75$ while at Magellan we used a
$1\arcsec$ slit \citep[][]{Barthetal2005,Xiaoetal2011}.  If the bulge
size is smaller than the slit size and the bulge is much fainter than
the disk component, then the observed $\sigma_{\ast}$ will be
contaminated by light from the disk.  Since $\sigma_{\ast}$ is a
luminosity-weighted measurement, the contamination will artificially
increase $\sigma_{\ast}$ in an edge-on system, and could slightly
lower it in a face-on system.  Here we evaluate whether
disk light contamination is important, and in what sense.  We divide
our sample into four bins based on bulge $r_e$: $60\%$ of the bulges have
$r_e <0\farcs37$, $22\%$ have $0\farcs37 < r_e < 0\farcs75$, 
$6\%$ have $0\farcs75 < r_e < 1\arcsec$, and the rest ($12\%$) have $r_e>1 \arcsec$.
Compared with our slit sizes, we conclude that disk
contamination can be significant.

As an additional test of the effect of disk contamination, we look 
for trends as a function of the galaxy inclination.  We use
our photometric fits to divide the sample with $\sigma_{\ast}$
measurements into three equal inclination bins 
containing 10 galaxies each.  Extrapolating the Faber-Jackson relation of
classical bulges defined by the sample from \cite{Gutekinetal2009} to
low luminosities, we calculate a predicted velocity dispersion
$\sigma_{\mathrm{predict}}$ from the observed magnitude.  We define
the average difference $\langle\sigma_{\mathrm{predict}} -
\sigma_{\ast}\rangle$. From edge-on galaxies to face-on galaxies, we
find $\langle\sigma_{\mathrm{predict}}-\sigma_{\ast}\rangle = 9,\ 20,\
32$ km s$^{-1}$ for the three bins respectively.  The galaxies with the 
highest inclination are those that suffer the most contamination.  On average, they 
also have the highest dispersions at fixed luminosity.  Thus, 
the disk contamination tends to boost the observed stellar velocity dispersions.

To avoid the complexity caused by disk contamination, we define the
subset of galaxies with stellar velocity dispersion measurements from
the ESI/MagE spectra and extended bulges with $r_e>0\farcs37$ to be
the clean sample. Most of the following analysis will be focused on
this restricted sample.

\subsubsection{Results}

In Figure \ref{FJrelation}, we show the relation between $I$-band
bulge magnitude and bulge velocity dispersion $\sigma_{\ast}$ in the
host galaxy. The clean restricted sample is labeled with open blue
circles filled with a black dot.  We compare our results
to the inactive early-type galaxies in \cite{Gutekinetal2009}.  We
choose this sample because it is a well-studied representative sample
of elliptical galaxies with very good measurements of luminosity and
velocity dispersion.  Our goal is simply to demarcate the region
occupied by elliptical galaxies in this plane, so that we can compare
with our sample distribution.  The $V$-band magnitudes reported in
G{\"u}ltekin et al. are shifted to the $I$-band assuming $V-I=1.34$ mag as
above. On average, our sample has smaller $\sigma_{\ast}$ at a fixed
bulge magnitude than the early-type galaxies.  The best fit relations
for G{\"u}ltekin et al.'s sample alone and the results with the clean
sample included are shown in Table \ref{fittingresult}. Our galaxies clearly 
have a different Faber-Jackson slope.  One way to
quantify the difference between our sample and classical bulges is to
calculate the average difference between the observed value
$\sigma_{\ast}$ and the predicted value $\sigma_{\mathrm{predict}}$,
as defined in Section \ref{sec:cleansample}.  For the clean
sample, we find
$\langle\sigma_{\mathrm{predict}}-\sigma_{\ast}\rangle=36 \pm 8$ km
s$^{-1}$, where the latter represents the error in the mean.

We can also quantify the difference in terms of bulge magnitude.  At
a fixed velocity dispersion $\sigma_{\ast}$, our sample galaxies have larger 
luminosities on average than do early type galaxies.
We can calculate the average difference $\langle M_{\rm
  predict}-M_I\rangle$, where $M_{\rm predict}$ is calculated based on
the observed velocity dispersion and the extrapolated Faber-Jackson
relation given by \cite{Gutekinetal2009}. For the clean sample, we get 
$\langle M_{\rm predict}-M_I\rangle=1.8$ mag with the error for 
the mean value to be $0.4$ mag. 

Not only is $\sigma_{\ast}$ for our sample significantly smaller at a
given bulge magnitude, but the scatter in the relation is also
significantly larger. In order to quantify the scatter, we fit the
log-linear relation $\log\sigma=\alpha+\beta M_I$ to the
data. Following the fitting method used in \cite{Tremaineetal2002} and
\cite{GreeneHo2006}, we include an intrinsic scatter such that the
reduced $\chi^2$ (equation 1 of \citealt{GreeneHo2006}) has a best-fit
value of one.  The sample from \cite{Gutekinetal2009} alone has an
intrinsic scatter of $0.064$ dex.  When we include our sample with
$\sigma_{\ast}$ from ESI/MagE spectra, the intrinsic scatter is
increased to $0.071$ dex. As shown in Figure \ref{FJrelation}, the
slope also gets steeper when our sample is included. We find
effectively the same results if we include the galaxies with
$\sigma_{\rm gas}$.
  
By comparing the Faber-Jackson relation for our sample and early-type
galaxies as shown in Figure \ref{FJrelation}, we find that our sample
has systematically smaller $\sigma_{\ast}$ than the predicted value
$\sigma_{\mathrm{predict}}$.  This offset, like those seen in the
photometric scalings above, is in the same sense as has been observed
for inactive pseudobulge samples \citep{KormendyIllingworth1983}.
Thus, in sum, the scaling relations are consistent with the fact that
$>70\%$ of our galaxies contain pseudobulges.

It is also interesting to compare the Faber-Jackson relation of our
sample with inactive spheroidal galaxies.
\cite{MatkovGuzma2005} and \cite{Codyetal2009} study the Faber-Jackson
relation for what they refer to as `dwarf early type' galaxies in the
Coma cluster with $R$-band magnitudes ranging from $-22.0$ to $-17.5$ mag
and $-20.7$ to $-15.6$ mag or $I$-band magnitudes of $\sim -23.2$ to
$-16.8$ mag \citep[][]{Fukugitaetal1995}.  According to their positions in
the fundamental plane \citep[e.g.,][]{GrahamGuzman2003}, these
galaxies obey our definition of spheroidal galaxies.
They are found to have a relation $L\propto
\sigma^2$, which is much flatter than the relation $L\propto \sigma^4$
for more luminous galaxies. As shown in Table
\ref{fittingresult}, when we fit our sample (galaxies with
$\sigma_{\ast}$ measurements) combined with that of G\"ultekin et al.,
we find $L\propto\sigma^3$.  Specifically, the 11 diskless galaxies 
lie very close to the spheroidal galaxies from \cite{MatkovGuzma2005}, 
and are systematically offset from the best fit relation of elliptical
galaxies. This is true even after we shift our sample to account for
different stellar populations, and confirms our conclusion from Figure
\ref{FPrelation2} that the diskless galaxies are actually
bright spheroidal galaxies.

\subsubsection{Uncertainties in the Faber-Jackson Analysis}

Conclusions for the contaminated galaxies are highly uncertain, since
we do not know the true $\sigma_{\ast}$ of these bulges.  Furthermore,
$\sigma_{\text{gas}}$ suffers from a variety of uncertainties. 
However, we note that our conclusions still hold
if we include the entire sample. Still, we focus on the clean sample 
because it is easiest to interpret.

Again, our sample likely contains younger stellar
populations than the early-type galaxies, which leads to offsets
in the Faber-Jackson relation. As we did for the fundamental plane
relation in $\S\ref{sec:FPrelation}$, we shift the bulge magnitudes by
one magnitude (Figure \ref{FJrelation}; note that if we adopt the
$M/L$ from Table 2 of \cite{GravesFaber2010} for both our sample and
classical bulges/elliptical galaxies, we only need to shift our sample
by $0.68$ mag). When we fit the relation $\log\sigma=\alpha+\beta M_I$
with the evolution correction applied, the intrinsic scatter remains
high at $0.070$, but the difference between the best-fit relations
including and excluding our sample is no longer significant. The
extent to which the offset between our sample and classical bulges is
intrinsic or can be explained by stellar population differences is
unclear at present.  This does not apply to the spheroidal galaxies, which 
remain offset even with the stellar population adjustment.

 \section{Discussion and Summary}
 \label{sec:summary}
 
We have looked at the host galaxies of 147 active galaxies selected
to have low-mass ($\mbh \lesssim 10^6 \msun$) BHs. Using
\emph{HST}/WFPC2, we perform detailed two-dimensional bulge, disk and
bar decompositions of the entire sample, to study the bulge
morphologies and structures of this unique sample. We find
that the sample is dominated by disk galaxies (only $7\%$ have no
disk) with small bulge components.  

We return to the questions we raised in the introduction.  Most host
galaxies of low-mass BHs have a bulge component.  Only seven objects
($5\%$) are consistent with being bulgeless galaxies. These are only
candidates for bulgeless active galaxies because of the fit
uncertainties, so BHs without any bulge component are apparently rare.
We do have to keep in mind that we are biased by optical selection,
which limits us to AGNs with high Eddington ratios in relatively
massive galaxies.  Multi-wavelength approaches are required to truly
determine the space density of AGNs in bulgeless galaxies.

Turning to those galaxies with bulges, the only remaining question is
whether they are classical or pseudobulges.  Based on the low
S{\'e}rsic indices, low bulge-to-total ratios, and the prevalence of
bars, rings, and nuclear spirals, we argue that the disk sample is
dominated by pseudobulge galaxies.  Consistent with this supposition,
we find that the fundamental plane of these galaxies is different from
both elliptical galaxies and spheroidal galaxies, but consistent with
observations of pseudobulge fundamental plane scalings.  In
particular, the galaxies are larger and have lower velocity
dispersions at a given luminosity than elliptical galaxies.  After we
account for their young stellar populations, the differences are even
more significant.  Our sample is also found to have a flatter
Faber-Jackson relation in the $L-\sigma$ plane.  Of the 147 galaxies,
we only find 13 galaxies with nearby companions that are candidates
for ongoing mergers, which is consistent with the fact that
pseudobulges evolve via secular processes.  In conclusion, the host
galaxies of low-mass BHs are different from classical bulges, and have
properties that are consistent with pseudobulges.  The 11 bulges
without extended disks do not scale as elliptical galaxies
either. Their fundamental plane scaling relations are systematically
offset, especially in the Faber-Jackson plane.  We have shown
definitively that a classical bulge is not a prerequisite to host a
supermassive BH.

We can also determine whether galaxies selected by their nuclear
activity differ in any way from field galaxies selected at the same
luminosity or mass. We have already seen that our disk galaxies have
comparable bar fractions and merger fractions as inactive galaxies.  We
now look at the distribution of morphological types for a more
complete sample of galaxies at similar luminosity.  We use the
morphology-dependent luminosity function from
\cite{Nakamuraetal2003}. For galaxies with $r^{\ast}$-band magnitudes
$M_r\approx-18$ mag, which corresponds to an $I$-band magnitude $M_I
\approx-19$ mag, the relative number of elliptical/S0, Sa/Sb spiral
galaxies and Sc/Sd spiral galaxies is found to be $1.00:4.10:0.73$.
Our sample has relatively more late-type spiral galaxies than field
galaxies selected by SDSS (Figure \ref{BTratio}).  The simplest
explanation for this observed difference is that we have a bias
towards spiral galaxies because they have the gas fuel needed to feed
a high-luminosity AGN.

Among spirals, inactive galaxies at the luminosity of our sample 
are dominated by pseudobulges. \cite{Weinzirletal2009} finds
that $\sim76\%$ of nearby high-mass spiral galaxies have low $n\le2$
bulges -- they are  pseudobulges. In a sample of 173 E-Sd galaxies,
\cite{FisherDrory2010} also finds that over $78\%$ of the identified bulges
are pseudobulges. This implies that our AGN-selected sample has very
similar bulge properties to non-AGN selected samples. 

Our results help elucidate the growth mechanisms of low-mass
BHs, as they are found preferentially in pseudobulges, which
are thought to be evolving secularly with a quiescent recent history
\citep[e.g.,][]{Kormendyetal2004}.  Most likely, these low-mass BHs
are not fueled by major mergers, which is also consistent with the
fact that only $9\%$ of our galaxies have close companions.  Rather,
the secular processes that build up the bulge may well fuel the AGN as
well.  In fact, it is possible that these BHs were formed with a mass
quite similar to their present mass (e.g.,
\citealt{PortegiesMcmillan2002,Koushiappasetal2004,Begelmanetal2006}).
If supermassive BHs indeed grow from low-mass BHs, some violent event
(e.g., a merger) is likely required to both dramatically change the
bulge properties and substantially grow the BH.

As already discussed in \cite{Greenetal2007} and \cite{Greenetal2010},
differences in the bulge properties of low-mass BHs lead to
differences in the $M_{\rm BH}$ --- $\sigma_{\ast}$ and $M_{\rm BH}$
-- $L_{\rm bulge}$ relations at low mass (if these relations still
exist at all). The $M_{\rm BH}$ -- $\sigma_{\ast}$ relation of this
sample is discussed in more detail in \cite{Xiaoetal2011} while the
$M_{\rm BH}$ --- $L_{\rm bulge}$ relation is discussed in a companion
paper \citep{Jiangetal2011}. In that paper, we find that the tight
scaling relations between BH mass and bulge luminosity found for
massive classical bulges do not exist for the pseudobulges that we
have shown are the hosts of low-mass BHs.

\section*{Acknowledgements}
Y.-F.J thanks Chien Peng for help in using GALFIT and Minjin Kim for
helpful discussions on fitting the images. We also thank the anonymous 
referee for valuable comments to improve the manuscript.
This work was supported by the Carnegie Institution for Science
and by NASA grant HST-GO-11130.01 from the Space Telescope Science
Institute, which is operated by AURA, Inc., under NASA contract
NAS5-26555. Research by A.J.B. is supported by NSF grant AST-0548198.

\begin{appendix}

\section{Tests of the Photometry}

In order to test for possible systematic biases in our fits, we estimate 
a nonparametric (model-independent) magnitude
for each galaxy. From the counts in the image we determine the
magnitude of the AGN and the galaxy as a whole. Although we cannot
decompose the galaxy into different components such as bulge and disk,
we derive a magnitude for the galaxy that is independent of any model
assumptions.  

The Tiny Tim PSF model described in \S\ref{sec:PSF} is used to
separate the AGN component from the host galaxy in the image. The only
assumption here is that the AGN component dominates the luminosity
within a small aperture at the center of the galaxy and that the AGN
component can be modeled by the PSF. We scale the AGN luminosity to
the flux within a two-pixel radius located at the center of the
galaxy.  For the reasons given in 
\cite{Jahnkeetal2004}, we do not scale the PSF model to a single pixel
at the center of the galaxy, because then we would be very sensitive
to centering errors. To determine the galaxy magnitude, we first 
subtract the sky from each pixel in the
object image. Second, we scale the PSF image so that the total
counts within an aperture located at the center of the PSF image are
the same as the total counts within this same aperture in the object
image. The magnitude of the scaled PSF model is taken to be the
nonparametric AGN magnitude.  Third, we subtract the scaled PSF model
from the object image.  The image with the AGN model subtracted
represents only the host galaxy, from which we measure a nonparametric
magnitude for the host galaxy as described below.

Typically, the galaxy does not fill the whole image and there are
usually other objects nearby. We thus measure the host galaxy
magnitude within a region of $700\times 700$ pixels in the subtracted
image and mask the contaminants.  There are 17 galaxies that may
extend beyond the PC chip.  In these cases, we create a composite
image with the three WF chips and measure the sky value from
these. Images from the four chips are combined with the command {\it
  wmosaic} in IRAF, which bins the image from the PC chip by a factor
$\sim2\times2$ so that the final image has a uniform pixel scale. The
boundaries between different chips are also taken care of
automatically by {\it wmosaic}. Images from other chips are reduced in
the same manner as the PC images (see Section \ref{Sec:Data}).  While
we may slightly underestimate the galaxy magnitudes, we do not include
regions from the WF chips in our nonparametric magnitude estimate.

The nonparametric magnitudes of the AGN components and the host
galaxies for all 147 images are given in Table \ref{parameters}. As
pointed out in \cite{Greenetal2008}, the host galaxies are generally
much brighter than the AGNs.  The median ratio between the light from
the AGN and the host galaxy is only $5\%$.  The difference between
nonparametric magnitudes and the magnitudes from our 
fits is always $<10\%$.

\begin{figure}
\centering
\includegraphics[width=1.0\hsize]{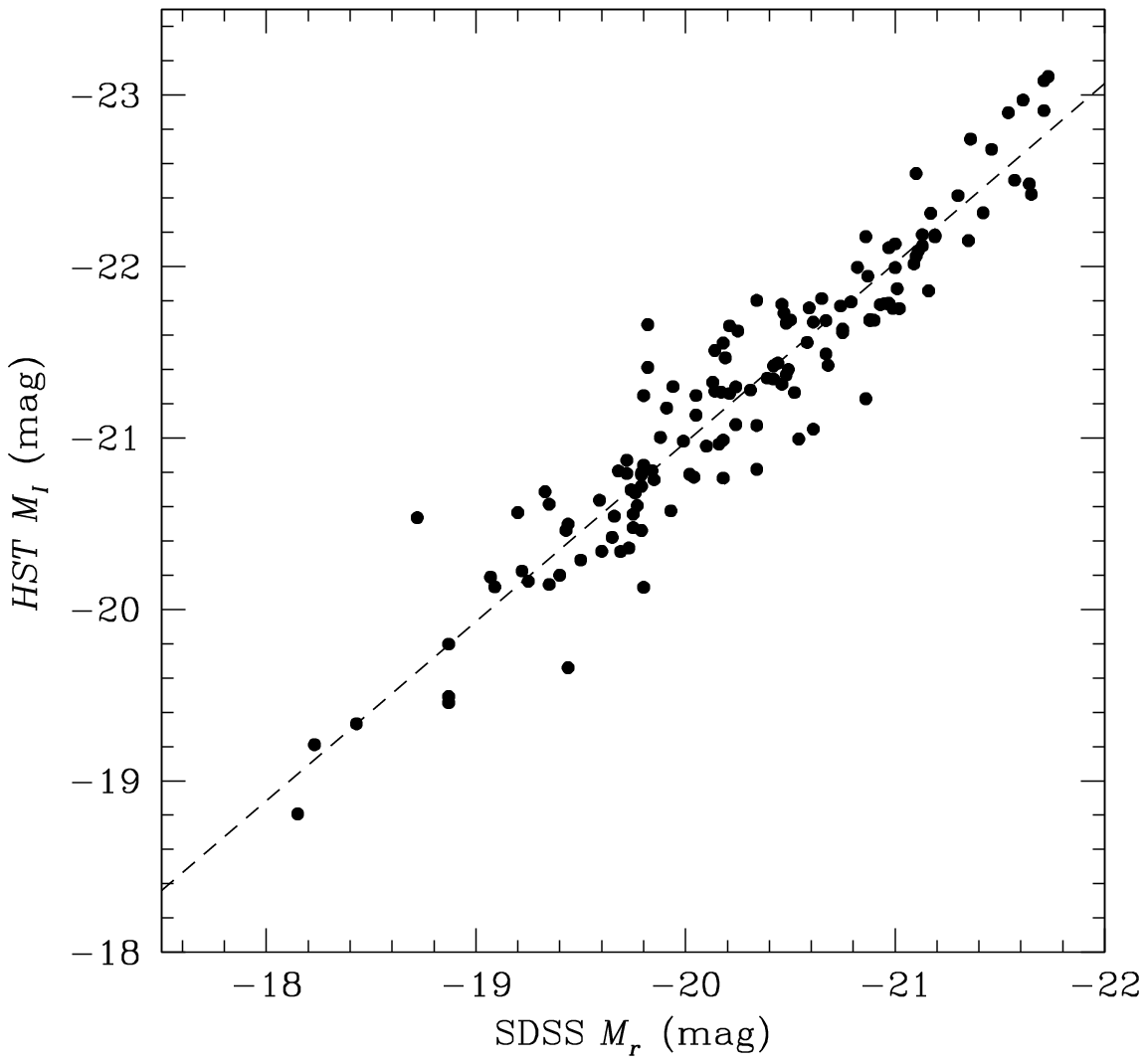}
\caption{Compare the $I$ band total magnitude from our best fitting 
model based on \emph{HST} images with $r$ band SDSS magnitudes.
}
 \label{checkmag}
\end{figure}

As an additional test of both our models and the sky levels in
particular, we also compare the $I$-band total magnitudes from our
models with the SDSS Petrosian $r$-band magnitudes as listed in Table
1 of \cite{Greenetal2007}. The comparison can be seen in Figure
\ref{checkmag}. We fit a linear relation
$M_I=\alpha(M_r-M_{r,0})+\beta$ to the two kinds of independent
magnitudes using the same fitting method described in Section
\ref{sec:FPrelation} and \cite{GreeneHo2006}. Here we assume
uncertainties in the magnitudes of $0.1$ mag (the median from our
fits).  We find $\alpha=1.05\pm0.03$, $\beta=-21.30\pm0.02$,
$M_{r,0}=-20.31$ and intrinsic scatter $e_0=0.22$. This means the
total magnitudes from our best fits are consistent with SDSS
magnitudes at the $2\sigma$ level, and suggests that we are not introducing 
major systematic errors through our treatment of the sky.

\end{appendix}

\end{document}